\documentclass{pasj00}

\begin{document}
\SetRunningHead{Hagihara et al. }{X-ray Spectroscopic Study toward the Galactic Bulge}
\Received{2011/02/25}
\Accepted{2011/08/22}

\title{An X-ray spectroscopic Study of the Hot Interstellar Medium Toward the Galactic Bulge}

\author{Toshishige \textsc{Hagihara}\altaffilmark{1}
  \thanks{Present Address is Cybozu Inc.,  1-4-14 Kouraku, Bunkyo, Tokyo 112-0004}, 
  Noriko Y.  \textsc{Yamasaki}\altaffilmark{1}, 
  Kazuhisa \textsc{Mitsuda} \altaffilmark{1},
  Yoh \textsc{Takei}\altaffilmark{1},
  Kazuhiro \textsc{Sakai}\altaffilmark{1},
  Yangsen \textsc{Yao}\altaffilmark{2},
  Q. Daniel \textsc{Wang}\altaffilmark{3},
  Dan \textsc{McCammon}\altaffilmark{4}
  }
  \email{yamasaki@astro.isas.jaxa.jp}
 \altaffiltext{1}{Institute of Space and Astronautical Science, Japan Aerospace Exploration Agency\\
3-1-1 Yoshinodai, Chuo, Sagamihara, 252-5210}
 \altaffiltext{2}{University of Colorado, CASA, 389 UCB, Boulder, CO 80309, USA}
 \altaffiltext{3}{Department of Astronomy, University of Massachusetts, Amherst, MA 01003, USA}
 \altaffiltext{4}{Department of Physics, University of Wisconsin, Madison, 1150 University Avenue, Madison, WI 53706, USA}


\maketitle

\begin{abstract}
We present a detailed spectroscopic study of the hot gas toward the Galactic bulge along the 4U 1820$-$303 
sight line by a combination analysis of  emission and absorption spectra.
In addition to the absorption lines of O\emissiontype{VII} K$_{\alpha}$, 
O\emissiontype{VII} K$_{\beta}$, O\emissiontype{VIII} K$_{\alpha}$ and   Ne\emissiontype{IX} K$_{\alpha}$
by Chandra LTGS as shown by previous works, Suzaku detected clearly the emission lines of 
O\emissiontype{VII}, O\emissiontype{VIII}, Ne\emissiontype{IX} and  Ne\emissiontype{X} from the 
vicinity.  We used  simplified plasma models with constant temperature and density. 
Evaluation of the background and foreground 
emission was performed carefully, including stellar X-ray contribution based on the recent X-ray 
observational results and stellar distribution simulator. If we assume that one plasma component exists 
in front of 4U1820$-$303 and the other one at the back, the obtained temperatures are
$T= 1.7\pm 0.2 \times 10^{6}$ K for  the front-side  plasma and 
$T=3.9^{+0.4}_{-0.3}\times 10^{6}$ K for the back-side. This scheme is consistent with a
hot and thick ISM disk  as suggested by the extragalactic source observations  and an X-ray bulge 
around the Galactic center.
\end{abstract}

\section{Introduction}
X-ray surveys,  such as the  ROSAT All Sky Survey (RASS),  show that  there is a large 
enhancement of the emission 
in the  Galactic center region expanding to $\pm$30$^{\circ}$ in longitude and latitude, which is called an X-ray
bulge.  
\citet{snowden97}  summed 3/4 keV (0.4 -- 1.2 keV) and 1.5 keV (0.7--2.0 keV) RASS data with 
10$^{\circ}$ wide bins centered on $\ell= 353^{\circ}$ and checked  the latitude profile of the surface 
brightness. They found that the intensity distribution of the enhancement in the $b>0^{\circ}$ region 
and the $b<0^{\circ}$ region significantly differed. They construct an isothermal ($T=10^{6.6}$ K) 
 cylinder plasma model with an exponential fall-off in density with height above the plane
  and showed that the enhancement for  $b<0^{\circ}$ can be well explained by this model. 
  The hot gas cylinder is located at the galactic center and its required radius is 5.6 kpc, 
  with an electron density of $3.5\times10^{-3}$ cm$^{-3}$ in the disk and a scale height of 
  1.9 kpc. They also suggest that in   the $b>0^{\circ}$ region, this model always produces 
  less intensity than the data and this implies an additional component,  such as Loop I. 
  The total luminosity of the plasma cylinder was $1.9\times10^{39}$ erg s$^{-1}$
  and the mass was $\sim 3 \times 10^{7} \MO$.
A molecular cloud of  $\sim10^{21}$ cm$^{-2}$ hydrogen column density is opaque to  soft X-rays 
lower than 0.5  keV. Shadowing observations using such molecular clouds located a few kpc away provide 
 important clues.
\citet{park97} observed the $(\ell, b)=(-10^{\circ}, 0^{\circ})$ direction with 
ROSAT,   where a  molecular cloud is located  $\sim$3 kpc away, and  showed an anti-correlation 
of the molecular cloud density and X-ray intensity. They compare the X-ray intensity in on-cloud and off-cloud directions 
to find that almost 40\% of the emission in both the 3/4 and 1.5 keV band comes from behind the cloud.
\citet{almy00} also carried	 out a shadowing analysis using  ROSAT PSPC data for  $(\ell,b)\sim(337^{\circ}, 4^{\circ})$, 
with an absorption column density assumed from IR (100 $\micron$) intensity, and revealed that 
more than 70 \% of the emission is due to X-ray sources behind the cloud. 

Recent studies  reveal that there are many kinds of X-ray emitters in our Galaxy. Not only the hot 
ISM in the Galactic disk and halo \citep{yao09, hagihara10, yoshino09}, but also CVs and normal stars can 
contribute to the unresolved emission in the Galaxy \citep{masui09}.
In order to distinguish between the diffuse plasma emission and the sum of faint point sources, 
measurements of the absorption line due to the plasma is essential, and combination study of the 
absorption and emission spectra can solve the density and scale  of the plasma, as shown by the 
observation of the hot halo around our Galaxy using extragalactic objects as background  
sources \citep{yao09, hagihara10}.

\begin{figure}[tbh]
\begin{center}
\FigureFile(80mm, 60mm){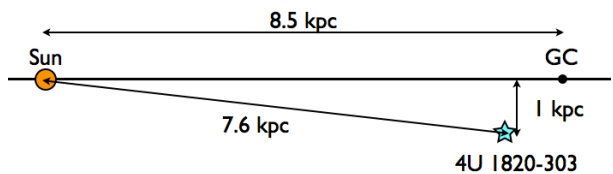}
\end{center}
\caption{Geometry of the sight  line toward 4U 1820$-$303 and the Galactic plane}
\label{fig:PositionalRelationships}
\end{figure}

Absorption observations toward the X-ray bulge  were carried out with an  X-ray grating detector. 
The X-ray binary 4U 1820$-$303 is located in the  globular cluster NGC 6624 at $(\ell,b)\sim(2.8^{\circ}, -7.9^{\circ})$, 
7.6$\pm$0.4 kpc away \citep{kuulkers03} as illustrated in figure \ref{fig:PositionalRelationships}. 
It was observed with the  high resolution grating onboard 
Chandra and absorption lines of   O\emissiontype{VII}, O\emissiontype{VIII} and Ne\emissiontype{VIII} were reported 
\citep{futamoto04, yao06}.
\citet{futamoto04} analyzed data using curve of growth and constrained the column density 
of each ion as $\log N_{\mathrm{O\emissiontype{VII}}}=16.2 - 16.7$, 
$\log N_{\mathrm{O\emissiontype{VIII}}} = 15.9 -16.5$ and $\log N_{\mathrm{Ne\emissiontype{IX}}} = 15.7 - 16.1$. 
The velocity dispersion ($v_{b}$) was derived using joint analysis of O\emissiontype{VII} K$\alpha$ and 
O\emissiontype{VII} K$\beta$ to $v_{b}> 200$ km s$^{-1}$,  and the column density of 
O\emissiontype{VIII}   and Ne\emissiontype{IX}  are derived assuming the same velocity dispersion 
constraint as O\emissiontype{VII}. Meanwhile \citet{yao06} used their absline absorption model 
on the same data and reported 
the column density as $\log N_{\mathrm{O\emissiontype{VII}}} = 16.3\pm 0.2$, $\log N_{\mathrm{O\emissiontype{VIII}}} = 16.4\pm0.2$ and 
$\log N_{\mathrm{Ne\emissiontype{IX}}} = 16.0\pm 0.1$ and the velocity dispersion was $255^{+114}_{-90}$ km s$^{-1}$. 

In this paper, we will use an emission spectrum  obtained by Suzaku for a region  $\sim 2^{\circ}$ away from 4U1820$-$303.
A combined analysis  
using  the   same method as used in the Galactic halo study with extragalactic objects
by \citet{yao09} and \citet{hagihara10}, in consideration with foreground diffuse and  discrete emission
 will give us further information about the density and scale of the hot gas in the
 Galactic bulge region. 
 
\section{Observation and Data Reduction}

\begin{table*}
\begin{center}
\caption{Observation logs}\label{tab:obslog}
\begin{tabular}{llllll}
\hline
Target name & $(\ell, b)$                             & ID &  Instrument                         &Obs.Start date & Exposure (ksec) \\ \hline
4U1820-30   & $(2.8^{\circ}, -7.9^{\circ})$ & 98 & Chandra HRC-S+LETG &2000-03-10  & 15.1  \\
 Bulge 3          &$(1.3^{\circ}, -7.5^{\circ})$  & 500001010& Suzaku XIS                       &2006-03-06 & 51.9 (32.2)$^{\dagger}$\\
\hline
\end{tabular}
\end{center}
\begin{footnotesize}
$^{\dagger}$ Value in parenthesis is the exposure time after data screening (see text)
\end{footnotesize}

\end{table*}
The observation logs of the Chandra and Suzaku data are summarized in table \ref{tab:obslog}.
4U1820$-$303 was observed  three times by Chandra,  but we used only the HRC-S + LETG data 
to obtain the best energy resolution. The dataset and data reduction process is basically the  same as 
in \citet{yao06}, but we
followed the procedure presented in \citet{yao09} to extract the first order spectra of the LETG.

Our Suzaku observations were taken with the CCD camera 
X-ray Imaging Spectrometer (XIS;
\cite{koyama07}, \cite{mitsuda07} ).
The XIS was set to the normal clocking mode  and the spaced-raw charge 
injection (SCI) was applied to the data during the observations. 
We used processing data version 2.2.7.18 for the two observations.
In this work, we  used only the spectrum obtained with XIS1.

We adopted the same data screening as in \citet{hagihara10}, i.e., 
standard data screening in addition to the exclusion of the 
thick atmospheric neutral oxygen column density in the line of sight \citep{smith07}
and of the Solar wind charge exchange (SWCX) \citep{fujimoto07}.
We found that the counting rate was constant as a function of the neutral oxygen  column
 density for the cleaned data. Thus, there is no significant neutral oxygen  emission 
 from the Earth's  atmosphere in the filtered data.
 The solar wind intensity  obtained with the Solar
Wind Electron Proton and Alpha Monitor (SWEPAM) aboard the 
{\it Advanced Composition Explorer} ({\it ACE }) was checked.  
We removed the time intervals
when the proton flux in the solar wind exceeded 
$4 \times 10^8$  cm$^{-2}$ s$^{-1}$  \citep{masui09}.
With this criterion, we discarded 19.7 ksec of the exposure and the net exposure time became 
32.2 ksec.

There are no obvious discrete X-ray sources in the X-ray images for the   0.4--1.0 keV energy range.
However, there would be many low luminosity sources in this direction, 
thus we applied  a wavelet analysis using CIAO:  wavdetect,  to detect such point sources. 
We found 23 point source candidates  in the fields and removed a circular region with 1' 
radius centered at the source position from the data. These candidates  can be caused by 
statistical fluctuations, and so we have to check them. We then 
compared the spectrum including the possible point sources  and the spectrum 
excluding them and found there is no significant difference.
To increase the photon counts and reduce the statistical error, we decided to use data 
including the point source candidates  in further analysis.

We constructed instrumental response files (rmfs) and effective area
files  (arf) by running the scripts {\it xisrmfgen} and {\it
xissimarfgen} \citep{ishisaki07}.
 To take into account the stray light coming from outside of the CCD fov, we
used a 20$'$ radius flat field as the input emission in calculating the
arf. We also included in the arf file the degradation of low energy
efficiency due to the contamination on the XIS optical blocking filter.
The versions of calibration files used here were
ae\_xi1\_quanteff\_20080504.fits, ae\_xi1\_rmfparam\_20080901.fits,
ae\_xi1\_makepi\_20080825.fits and ae\_xi1\_contami\_20071224.fits.
We estimated the non-X-ray-background from the night Earth database
using the method described in \citet{tawa08}.

\section{Analysis and Results}

\subsection{Absorption spectrum analysis}
We first calculated  the equivalent widths (EWs). We constructed a model with a power-law and 
narrow gaussians and fitted the spectra in narrow ranges as shown in table \ref{tab:abs_ew}, 
and obtained EWs for  each line using the eqwidth tool in Xspec. 
The errors in the equivalent widths are obtained as following; 
1) we calculated the 90\% error range of the normalization of the gaussian function. 
2) Next, we calculated the maximum error of the EW ($EW_{\mathrm{max}}$) using normalization values of 
the best fit ($N_{\mathrm{best}}$) and maximum ($N_{\mathrm {max}}$) 
as $EW_{\mathrm{max}} = EW/N_{\mathrm {best}} \times N_{\mathrm{max}}$.
Comparing the previous work, our obtained value for the O\emissiontype{VII}  K$\alpha$ ($0.65^{+0.19}_{-0.14}$ eV )
is  smaller than those in \citet{futamoto04}  ($1.19^{+0.47}_{-0.30}$ eV) and \citet{yao06}
($1.06^{+0.34}_{-0.27}$ eV).  
We confirmed  that this difference comes from the modeling  of 
the continuum spectrum, with higher order diffraction photons \citep{yao06}.
We used only the 1st order spectrum of the HRC observation, and modeled the continuum spectrum in a narrow energy range, to obtain reliable values.
\begin{table*}
\caption{Equivalent widths of absorption lines in 4U1820-303 spectrum}\label{tab:abs_ew}
\begin{center}
\begin{tabular}{lccccc}
\hline
       & O\emissiontype{VII} K$\alpha$ & O\emissiontype{VIII} K$\alpha$ &O\emissiontype{VII} K$\beta$  &Ne\emissiontype{IX} K$\alpha$ &Ne\emissiontype{X} K$\alpha$\\
  \hline
  Centroid (keV) &  0.5734 & 0.6537 & 0.6663 & 0.9223 & 1.022 \\
  EW (eV)  & $0.65^{+0.19}_{-0.14}$ & $0.69^{+0.17}_{-0.20}$ &$0.19^{+0.16}_{-0.16}$ &$0.52^{+0.24}_{-0.22}$ &$<0.16 $ \\
  EW (m\AA)& $24.4^{+7.1}_{-5.2}$& $19.9^{+4.9}_{-5.8}$&  $5.3^{+4.4}_{-4.4}$& $7.5^{+3.4}_{-3.2}$& $<1.9$ \\
  Fitting Range(\AA)& 20-22& 18--20 & 18--20 & 12--14 & 11.4-12.5 \\ 
  \hline
\end{tabular}
\end{center}
\end{table*}%

To obtain the column density of ions and to estimate the total column density of the 
absorbing material, we  fit all the lines simultaneously. We first fitted the continuum 
between 0.54 and 1.1 keV with a continuum model of a blackbody and a broken power-law. 
Foreground absorption by the neutral ISM with  solar metal abundances by \citet{anders89}  is assumed. 
This model returned a minimum $\chi^{2}$/dof= 858.1/577 and 
the best-fit column density of the neutral ISM $N_{\mathrm {H}}$ is $1.9\times10^{21}$cm$^{-2}$.
This value is consistent with $N_{\mathrm {H}}$  estimated from 
by the 100 $\mu$m  intensity ($2.3\times10^{21}$cm$^{-2}$)
 and reported in \citet{yao06}.

This poor fit might be caused by residuals at  the metal edge due to uncertainties in 
the Chandra response matrices.
\citet{nicastro05} set the metal abundance of the neutral absorption
 material to be free to deal with this problem. As we followed their method, it decreases the residuals 
 ($\chi^{2}$/dof=761.1/560) with the virtual absence of Ne and Na, and an over-abundance of 
 C, N and Mg by factors of 3, 3 and 26 respectively. These values are due to the uncertainties
  in calibration, rather than true ISM metallicity \citep{nicastro05}. 
  Though the $\chi^{2}$ is still not good, it is difficult to compensate the calibration 
  uncertainties further and we use this model to describe the continuum.
  
\begin{figure}[htb]
\begin{center}
\FigureFile(0.45\textwidth, 80mm){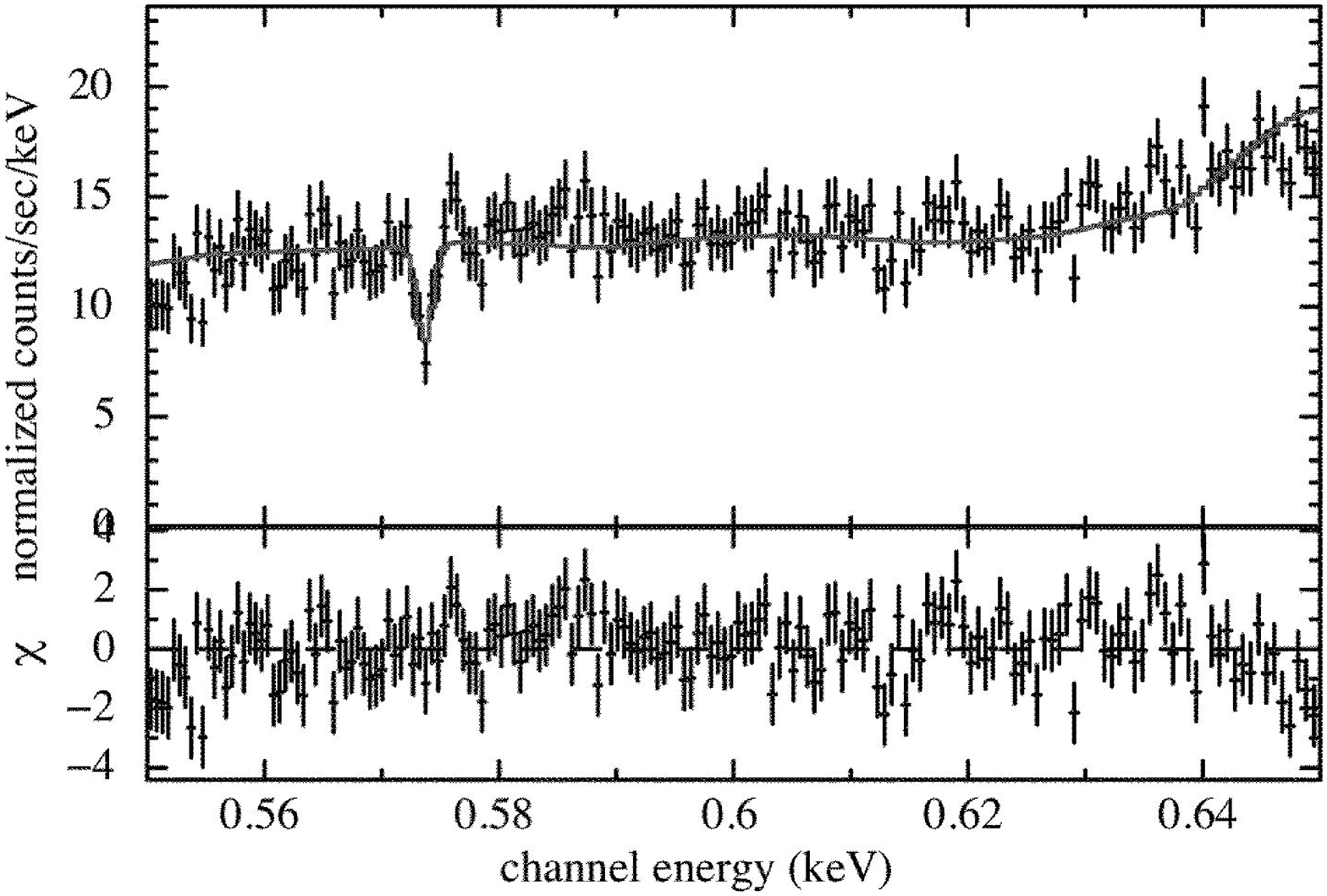}
\FigureFile(0.45\textwidth, 80mm){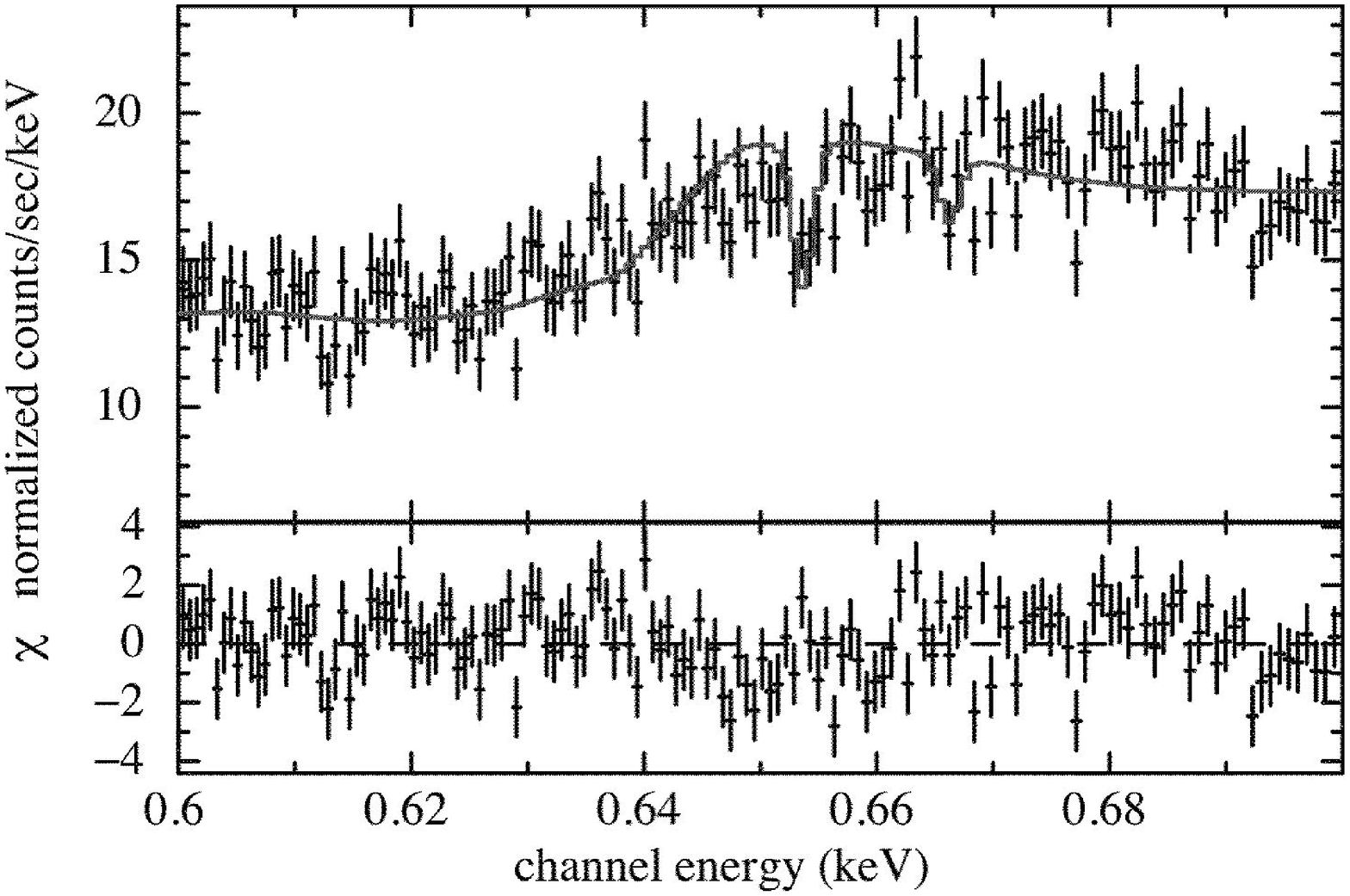}
\FigureFile(0.45\textwidth, 80mm){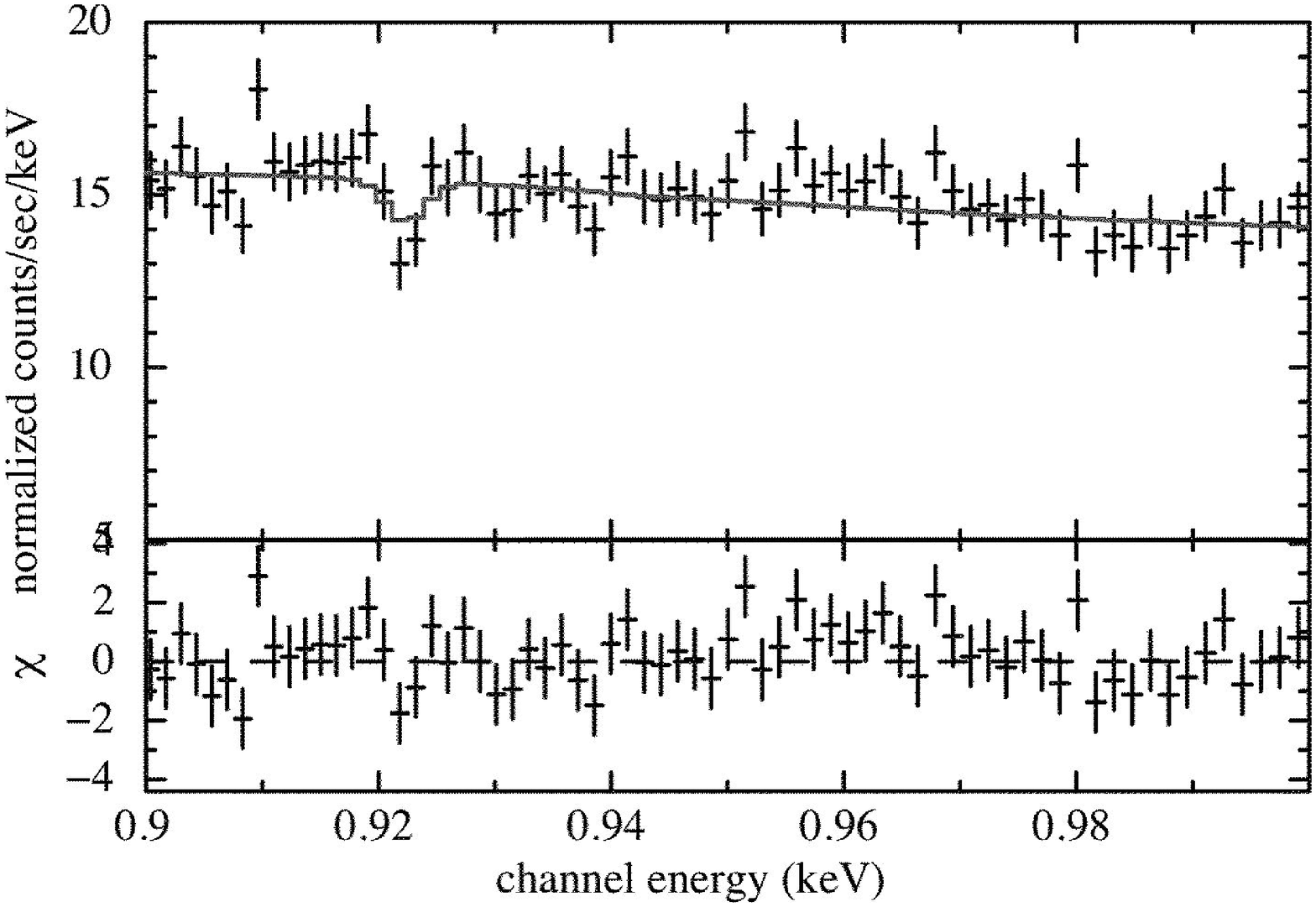}

\caption{O and Ne absorption lines in the 4U 1820$-$303 spectrum. 
The solid lines are model A4.}
\label{fig:AbsLine}
\end{center}
\end{figure}

  We next added two absorption lines representing O\emissiontype{VII} K$\alpha$ and O\emissiontype{VII}
   K$\beta$  using the absem model,  which is the same as in \citet{hagihara10}. 
   The column density and velocity dispersion of these lines are linked because
    both lines  originate from  the same ionization state of oxygen (model A1). 
    The results are given in table \ref{tab:abs_column}. 
     We next added absorption lines representing O\emissiontype{VIII} K$\alpha$
    and Ne\emissiontype{IX} K$\alpha$ step by step (model A2, A3). 
    In each step, the velocity dispersion of lines are linked together. 
    The results maintain consistency, thus in further analysis the velocity dispersion
     is always linked together. For the next step, we evaluated the hot phase hydrogen 
     column density assuming solar abundances of O and Ne (model A4) (see figure \ref{fig:AbsLine}).
     We then  let the Ne abundance vary and fitted the spectrum (model A4$'$). Please note that we fixed 
     the abundance of O to the solar value, because the contribution of continuum emission is 
     small and it is hard to determine the absolute metallically.      
     Though the best fit hydrogen column density become smaller and the 
     Ne abundance is $2.2^{+1.8}_{-1.3}$, all the parameters are consistent with those of the A4 model.
     We will use this A4 model for further combined analysis, and evaluate the Ne/O abundance effects 
     if necessary.
There might be intrinsic absorbing material around the 4U1820$-$303. \citet{futamoto04} 
 discuss the possibility of intrinsic absorption using a photo-ionization simulator and conclude
  that the size of the binary system and the luminosity cannot explain the O\emissiontype{VII} 
   ionization fraction estimated by the column density ratio. Thus we consider that all of these absorption
    lines  originate in the  hot ISM in the further analysis.
    
\begin{table*}
\caption{Parameters of the hot ISM absorbing the 4U1820$-$303 spectrum}
\label{tab:abs_column}
\begin{center}
\begin{tabular}{lcccccccc}
\hline
Model & $v_{b}$&\multicolumn{4}{c}{$\log N$} & Abundance & $\log T$ & $\chi^{2}$/dof \\ 
            & (km s$^{-1}$) &\multicolumn{4}{c}{(cm$^{-2}$)} &   & (K)  & \\
             &                & O\emissiontype{VII} &  O\emissiontype{VIII} & Ne\emissiontype{IX} & ${\rm H_{HOT}}$ & Ne/O&  $T_{1}$ &  \\
 \hline
 A1 & $163^{+219}_{-77}$ & $16.17^{+0.33}_{-0.21}$ & $\cdots$    &  $\cdots$ & $\cdots$ &  $\cdots$& $\cdots$ &708.4/573 \\
 A2 & $239^{+301}_{-102}$ &$16.10^{+0.37}_{-0.16}$ &$15.90^{+0.37}_{-0.17}$ &$\cdots$ &$\cdots$ &$\cdots$ &$\cdots$ & 671.2/571 \\
 A3 & $191^{+254}_{-101}$ &$16.14^{+0.35}_{-0.19}$ &$15.92^{+0.42}_{-0.19}$ &$15.87^{+0.30}_{-0.33}$ &$\cdots$ &$\cdots$ & $\cdots$ & 759.9/569 \\
 A4 & $112^{+114}_{-26}$ &$\cdots$   &$\cdots$    &$\cdots$   & $19.66^{+0.16}_{-0.16}$ &$\cdots$  & $6.27^{+0.13}_{-0.07}$ & 662.1/570 \\
 A4$'$ &  $187^{+254}_{-98}$ & $\cdots$  &$\cdots$    &$\cdots$     &  $19.46^{+0.34}_{-0.15}$ &  $2.2^{+1.8}_{-1.3}$ & $6.28^{+0.06}_{-0.06}$ & 659.8/569 \\
 \hline 
\end{tabular}
\end{center}
\label{default}
\end{table*}%

\subsection{Emission spectrum analysis}
\subsubsection{Modeling of the contribution from foreground emission and stars}
\begin{figure}[htb]
\begin{center}

\FigureFile(0.4\textwidth, 50mm){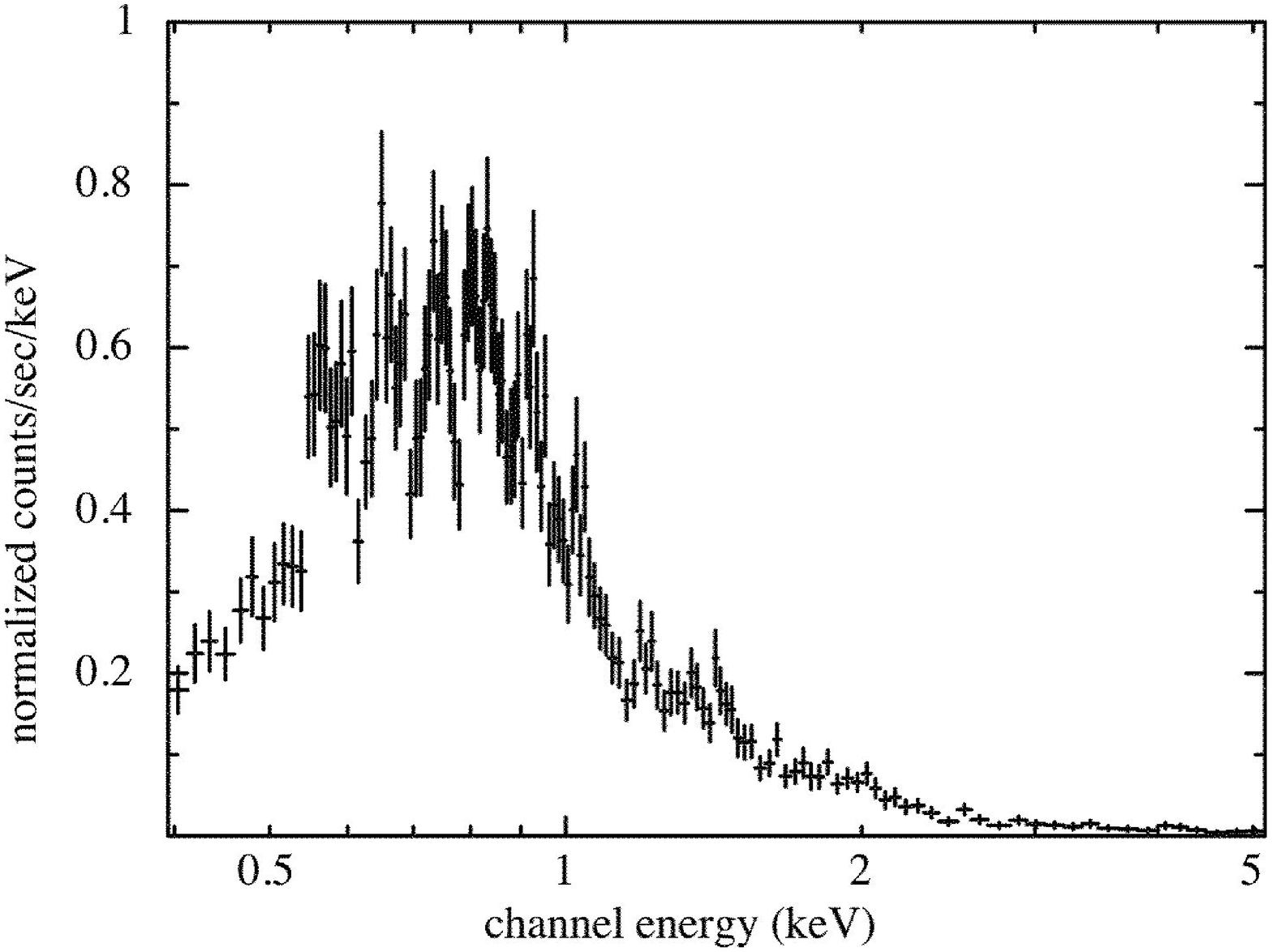}
\FigureFile(0.45\textwidth, 100mm){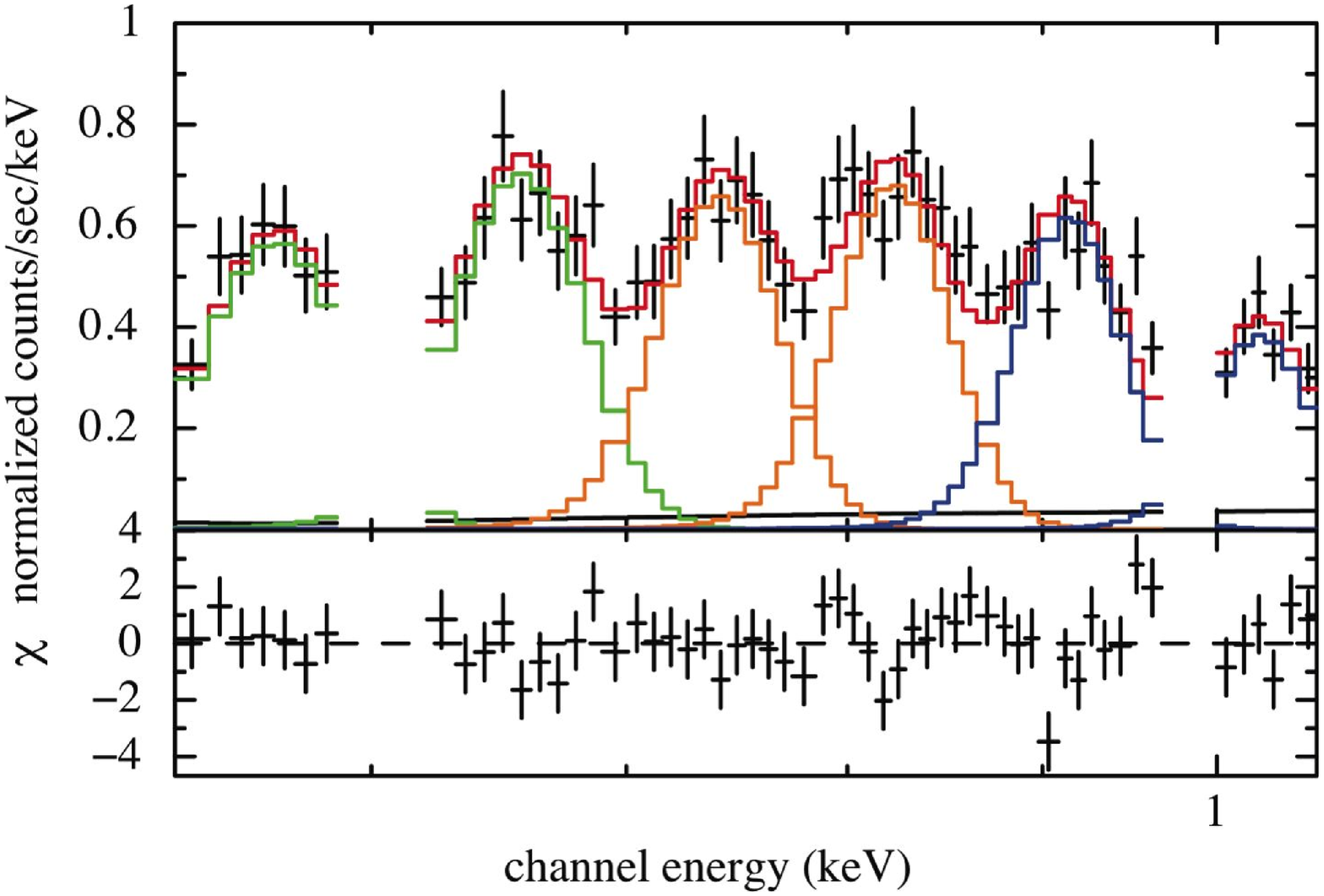}

\end{center}
\caption{Spectrum obtained in the galactic bulge region (left) and evaluation 
of the line intensities by Gaussians.(right)
}
\label{fig:EmissionSpectrum}
\end{figure}

The emission spectrum from the 4U1820-303 vicinity obtained by Suzaku is shown in
figure \ref{fig:EmissionSpectrum}. The emission below 2 keV is very bright,  and 
we can easily distinguish the emission lines of O\emissiontype{VII}, O\emissiontype{VIII},
Ne\emissiontype{IX}, and Ne\emissiontype{X}. We evaluated the intensity of these lines by 
a simple power-law and Gaussians fitting model (model E1) and the results are summarized in table \ref{tab:LineSB}. 
Hereafter, the intensity of the line emission is shown in LU (Line Units), which corresponds to 
photons s$^{-1}$ cm$^{-2}$str$^{-1}$.
These line ratios are hard to reproduce with a single -temperature plasma.  Thus we will study several components
that can contribute to  the emission in this region. We will consider  the diffuse emission from the Solar  neighborhood, 
Loop I, and contribution from stars in the Galaxy,   step by step.

\begin{table}
\caption{Apparent  surface brightness of each line. (model E1)}
 \label{tab:LineSB}
\begin{center}
\begin{tabular}{cccc} \hline \hline
O\emissiontype{VII} &O\emissiontype{VIII}  &Ne\emissiontype{IX}&  Ne\emissiontype{X}\\ 
(LU) &(LU) &(LU) & (LU)\\\hline
$23.45^{+3.21}_{-1.71}$ & $18.96^{+0.87}_{-2.25}$ &
                 $10.49^{+0.49}_{-1.22}$ & $5.99^{+0.76}_{-0.58}$ \\ \hline
\end{tabular}
\end{center}
\end{table}

As shown by the Suzaku shadowing observations of  MBM-12 molecular clouds  \citep{smith07}
and the evaluation of the soft X-ray background spectra of 14 fields \citep{yoshino09}, 
emission around our Solar neighborhood due to Solar wind charge exchange and/or
the Local hot bubble is  present.  Most of the emission from the Solar neighborhood might 
be below 0.5 keV, except for $\sim$ 2 LU of O\emissiontype{VII} emission line.
The apparent spectrum above 0.5 keV can be reproduced by 
an optically thin thermal plasma emission in 
collisional ionization equilibrium (CIE)  with $k{\rm}T\sim 0.1 $ keV \citep{yoshino09}. 
As the uncertainty of the O\emissiontype{VII} line intensity is $\sim 1$ LU, 
we adopt these values as the foreground emission.

The sight line toward 4U 1820$-$303 would go through the Loop I,  a large structure 
seen in the radio waveband. This structure is considered to be  
an old SNR ($10^{6}$ years) located in Sco-Cen OB association.
\citet{egger95} modeled this SNR based on RASS data.
 The density 
and temperature in the cavity are
$2.5\times 10 ^{-3}$ cm$^{-3}$ and $4.6 \times 10 ^6$ K.
We then assumed  the column density and line intensity using these parameters 
as shown in table \ref{tab:LoopIContribution}.
After  comparing these values with those in table \ref{tab:abs_column} and \ref{tab:LineSB}, 
we will neglect the effect of Loop-I in the absorption line analysis, but take it into account 
in the emission line analysis. 

\begin{table*}
\caption{Estimated contribution from the Solar neighborhood and  Loop I }
\label{tab:LoopIContribution}
\begin{center}
\begin{tabular}{cccccccccc} \hline  \hline 
Model &  \multicolumn{4}{c}{Line Intensity }
 &
 \multicolumn{4}{c}{ Column Density }\\
&\multicolumn{4}{c}{(LU)}  &
 \multicolumn{4}{c}{ ($\log N$/(cm$^{-2}$)) } 
 \\
 &O\emissiontype{VII} & O\emissiontype{VIII} & Ne\emissiontype{IX} & Ne\emissiontype{X} &O\emissiontype{VII} & O\emissiontype{VIII} & Ne\emissiontype{IX} & Ne\emissiontype{X}  \\ \hline
 Solar neighborhood  & 2 & 0 & 0& 0& $\cdots$&  $\cdots$& $\cdots$ & $\cdots$ \\
Loop I  &  1.83 & 2.10 & 0.31 & 0.04 & 14.3 & 14.8 & 14.3 & 13.8 \\ 
\hline
\end{tabular}
\end{center}
\end{table*}

Recently, the contribution of stars in the soft X-ray background was studied using 
Chandra, XMM-Newton \citep{Lopez07} and Suzaku \citep{masui09}.
The typical luminosity of a stellar corona is as small as $10^{29}$ erg s$^{-1}$, but 
 the number density is large in the Galactic plane and bulge direction. We 
estimated the contribution from stars by the following steps. 
First, we estimated the number of stars in the field-of-view, using a stellar population model.
Secondly, we calculated the X-ray flux and spectra based on the currently available 
observational properties. 
 
We used TRILEGAL simulator ver 1.4 \footnote{http://stev.oapd.inaf.it/cgi-bin/trilegal}
\citep{girardi05} to estimate the main sequence stellar 
distribution in the observing cone.  This simulator synthesizes a stellar population 
for a given Galaxy field. \citet{vanhollebeke09} used this simulator and compared the results with 
Two Micron All Sky Survey (2MASS) \citep{skrutskie06} and Optical
Gravitational Lensing Experiment (OGLE) \citep{udalski97} observational data in 11 directions.
There is a discrepancy of about 20\% between the model and data, especially in the number of low 
luminosity stars.
We used their best model parameters, to create the mock population, and 
counted the stars of several spectral types, and ages as shown in table \ref{tab:TRILEGAL}.
Clearly, there are lots of A, F, G, K, and M stars in the  Suzaku  field of view.
Note that we count the stars in a radius of 20$'$ in the   Suzaku  sight line, and estimated the 
observed flux on the detector via the arf.
\citet{kuntz01} estimated the X-ray luminosity of these stars based on the  ROSAT data, 
and \citet{RocksD} compiled the X-ray spectra after \citet{Lopez07} which uses XMM-Newton observation. 
We used  the values shown in table \ref{tab:StarParameter} as a template of each spectral type stars.
In \citet{Lopez07}, about half of the K and M stars are represented by the two-temperature coronal model, 
and the other half is described by a single -temperature coronal model.
We added the flux from the mock star distribution with these X-ray luminosity and spectra, by 8 thin thermal 
plasmas as an empirical mock-up spectrum.
The RS CVn  type binaries are also bright in X-rays, with a typical X-ray luminosity $>10^{30}$ erg s$^{-1}$. 
We adopt a simple exponential disk model by  \citet{Ottmann92}, and found that the 
estimated numbers in a circle of 20$'$ is about $\sim$ 980.  
 
The HI column density toward the observing direction by 21 cm radio observation is 
$1.36\times 10^{21}$ cm$^{-2}$ \citep{kalberla05}. The HI +H$_{2}$ column derived by IRAS 100 $\mu$m 
intensity with a conversion formula by \citet{Snowden93} is $1.42\times 10^{21}$ cm$^{-2}$.
We also calculated the hydrogen column as a function of the distance from the Sun 
based on the global Galactic model by \citet{Ferriere98}, and found that $\sim$ 80\% of the
absorption material is located within 2 kpc from us.
Thus, as a crude assumption, we do not apply absorption by the neutral ISM to the 
SWCX+LHB component, Loop I, and stars and RS CVns within 2 kpc, but we  do apply the 
absorption of  $1.42\times 10^{21}$ cm$^{-2}$ for stars and RS CVns  beyond  2 kpc.

\begin{table*}
\caption{Criteria and number of stars in the direction of   Suzaku observation 
within a radius of 20' for each spectral type}
\label{tab:TRILEGAL}
\begin{center}
\begin{tabular}{cccrrrr} \hline\hline
spectral type & (B-V)$_0$ & $M_V$ &0-0.15 Gyr& 0.15-1 Gyr&1-10 Gyr &$>$10 Gyr\\
  \hline
O+B & $<$-0.01 &$\cdots$&8   &6 & 7 & 0 \\
A &-0.01-0.3 & $\cdots$ & 6  &58  &12 & 90  \\ 
F &0.3-0.6 & 2-8  &12  &78   &35212   & 41152 \\
G &0.6-0.8 & 2-10 &6  &58 & 97672 &  29270 \\
K & $>$0.8 & $<$8 & 41  &240  & 454739 &  72546 \\
M(early) &$>$0.8 & 8-15 &133&1054 &1851091 & 407848 \\
M(late) &$>$0.8 & $>$15 &0 &0  &35 & 0 \\
\hline
\end{tabular}
\end{center}
\end{table*}

\begin{table*}
\caption{Stellar type parameters used in stellar emission estimation. The empirical spectral 
model are summarized by  \citet{RocksD} after \citet{Lopez07} 
and luminosities estimated by \citet{kuntz01}.
RS-CVn type binary parameters after \citet{Ottmann92} are also shown.}
\label{tab:StarParameter}

\begin{tabular}{lcccccccccc} \hline\hline
Spectral  & Single  & \multicolumn{3}{c}{Two temperature model} &   Abundance & \multicolumn{3}{c}{$\log L_{\rm X}$(erg s$^{-1}$)$^{\dagger}$} \\
type & temperature & & & & & \multicolumn{3}{c}{Age} \\
 & $kT$ (keV) &$kT_{\rm hot}$ (keV) &$kT_{\rm cool}$ (keV) &EM$_{\rm cool}$/ EM$_{\rm hot}$  &$Z_{\odot}$ &0--0.15 Gyr& 0.15--1 Gyr& 1--10 Gyr\\ \hline
F &0.58 &$\cdots$ &$\cdots$&$\cdots$  & 0.5&29.51 & 29.02&28.13\\ 
G &0.67 &$\cdots$ &$\cdots$ &$\cdots$  & 0.3 &29.91& 28.82&27.89\\
K &0.83 &1.17 &0.32 &1.97  &0.2 &29.66& 28.52&27.89\\ 
M(early) &0.90 &0.80 &0.27 &2.02  &0.1 &29.31&28.49&27.54\\ 
RS-CVn  & N/A & 2.59 & 0.17 & 0.22 & 1.0 &\multicolumn{3}{c}{30.75$^{\ddagger}$}\\ 
\hline
\end{tabular}
$\dagger$ Luminosity in 0.1--2.4 keV\\
$\ddagger$  Luminosity in 0.4-4.0  keV independent of age
\end{table*}

\subsubsection{Hot ISM emission }
We then try to represent the observed energy spectrum with contributions from stars and RS CVns
and the Cosmic X-ray Background (CXB), which is modeled as an absorbed  power-law with a 
photon index of 1.4 and its  normalization of about 10 photon sec$^{-1}$ cm$^{-2}$ str $^{-1}$ keV$^{-1}$
at  1 keV \citep{hasinger93}. This model only reproduces about 1/10 of the emission below 1 keV. 
Even though we allowed the normalization of every stellar component to vary, it was impossible to 
exhibit enough O\emissiontype{VII} line because the temperature of the stellar components are all high.

\begin{figure}
\begin{center}
\FigureFile(0.5\textwidth, 80mm){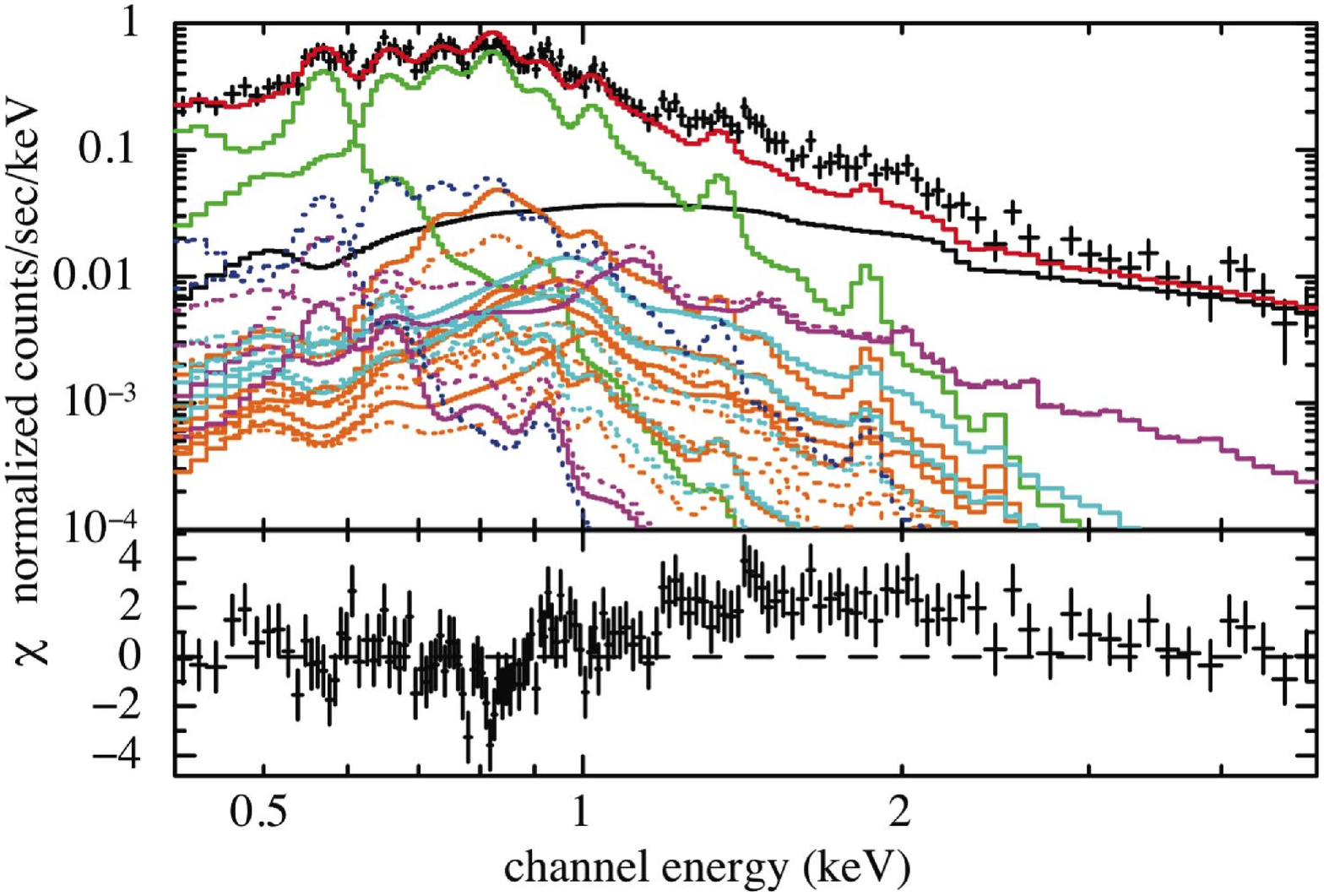}
\caption{Spectrum and model (E1). The dotted lines indicate foreground
 (unabsorbed) components and the solid lines indicate background (fully
 absorbed) components. hot ISM (green), CXB (black),  SWCX+LHB, 
 and Loop I (blue), stars except for M stars (orange), M stars (light blue),
and RS CVns (magenta).}\label{fig:AllFixed}
\end{center}

\end{figure}

We next added a hot ISM component to reproduce the spectrum.
We tried a one temperature hot ISM model with fixed and free (N, Ne and Fe) abundance  and 
obtained poor $\chi^2$/dof value of 578.62/133 and 366.20/130 respectively.
This poor fit is caused by the spectrum where O\emissiontype{VII} and
Ne\emissiontype{X} lines both exist  and it is  difficult for a single temperature
plasma to reproduce these lines simultaneously.
We then gave up the one temperature hot ISM model and tried a two temperature
model with fixed abundance (model E1).
Though the $\chi^2$ was improved to 358.17/131, there
are still significant residuals between the model and spectrum as shown in  figure  \ref{fig:AllFixed}.
It is obvious that a continuum-like component is needed to reduce the
residuals between 1 and 3 keV, thus we tried to free  the CXB parameters.
Though this model reduced the residuals and improved the  $\chi^2$/dof to 121.85/129,
 the best fit values of photon index is 2.4$^{+0.1}_{-0.1}$ and the
 normalization is 41.5$^{+3.6}_{-7.6}$ photon 
 sec$^{-1}$ cm$^{-2}$ str $^{-1}$ keV$^{-1}$ at 1 keV. The flux is about 4 times larger than 
 the nominal value \citep{hasinger93} and is not 
 reasonable for the CXB even though its fluctuation is taken into account.

As mentioned before, there are large uncertainties in the number
densities, luminosities , and the energy spectrum of the background low luminosity stars and binaries.
For this reason, we tried  the following two models.
\begin{itemize}
 \item Model E2: Assuming an underestimation of the stellar contribution,
	   the normalization of the background 
	   stars and RS CVns are set to be free.
 \item Model E3: Assuming an unknown thermal component in the bulge,  
	   one high temperature thin thermal plasma emission model is added.
\end{itemize} 

In the E2  model, we set the normalization of stars to be free step by step and 
investigated   the residuals and the normalization
 and estimated the effect of star contributions on the hot ISM properties.
To compensate for the residuals around 2 keV, we needed an additional $\sim$ 2keV
component. One plausible component to produce such hard emission is 
RS CVn binaries of which we modeled a spectrum  with a two temperature
plasma of 2.59 keV and 0.17 keV.
Moreover it is not well understood how many stars make binaries 
and there is uncertainty in the distribution of the RS CVn binaries.
Thus we first set the normalization of 
background binary components to be free 
(Model E2-b: The emission model with background binary normalization is set to be free).
Though we found that this model certainly reduced the residuals around 2 keV,
we also found that this model caused  significant residuals around 1 keV and above 2.5 keV.
Emission from plasma of $kT=$ 2.59 keV is a little  too high to
compensate for the residual.

The M type stars exhibit coronal emission whose temperature is empirically known to be  $kT =0.5\sim 1.2$ keV,
and characterized by complex emission with Ne, and Fe-L lines around 0.9 keV, 
\citep{Lopez07, Sciortino99}
  and their low luminosity could cause
large uncertainties in the number of stars within  the FOV.
For the next step, we fixed the binary normalization to the simulated value and set the 
normalization of the M type star
to be free  (Model E2-m: Emission model with background M type star normalization is set
to be free).
This model also could not explain the entire  spectrum.
However, residuals caused by these two models (E2-b and E2-m) are
complementary.

We thus  tried to free the normalization of binary and M stars
simultaneously (model E2-mb).
This model fit the data with $\chi^2$/dof of 144.37/131.
The normalization ratios to the simulated values are 6.57$^{+1.17}_{ -2.14}$
and 5.45$^{+1.27}_{-2.20}$ for M stars and binaries respectively.
It is difficult to confirm if these values are correct or not,  so 
we accept this value at present. 

We then set the normalization of the K type star and F and G type star to be free
sequentially (model E2-mkb, E2-mkfgb). 
This modification  caused  no significant changes to the fitting results and
the contribution of K, F and G stars  vanished.
Finally we linked the normalization of  all star components and 
set them to be free (model E2-(mkgf)b).
The  ratios of the obtained normalization to the simulated values are 2.79$^{+0.74}_{ -0.78}$
and 6.22$^{+1.35}_{-1.35}$ for all spectral type stars and binaries
respectively.
$\chi^2$/dof is 151.29/129 and it is hard to explain the whole spectra
with this model.

The normalizations and temperatures of the cooler and hotter hot ISM are
consistent with each other in the three  models (E2-mb, E2-mkb, E2-mkgfb).
From these results, the normalization of the stars is 
of little effect on the hot ISM temperature.

\begin{table*}
 \caption{Fitting results with model E2 (see text for details)}
\label{tb:FittingResultsE2}
\footnotesize
 \begin{center}
\begin{tabular}{lccccccccccc} \hline \hline
Model& Foreground$^{\dagger}$  & \multicolumn{4}{c}{ISM} &
 \multicolumn{4}{c}{Background$^{\dagger}$ } & $\chi^2$/dof\\ \cline{3-6} 
& &  \multicolumn{2}{c}{cool} &\multicolumn{2}{c}{hot} & \multicolumn{2}{c}{Stars} &
 RS CVn  & CXB & \\ 
& & $\log T$ (K) & Norm$^{a}$ & $\log T$  (K) &
 Norm$^{a}$ & Type$^{b}$ & Ratio$^{c}$ &Ratio$^{c}$  &\\ 
 & &  [$kT$ (keV)] & &  [$kT$ (keV)]&  &  &  &  &\\ 
 \hline 
E2-b &fixed &$6.009^{+0.047}_{-0.083}$ &$215.4^{+319.9}_{-87.6}$
			 &$6.571^{+0.025}_{-0.022}$ &$26.3^{+2.7}_{-2.6}$ &- & fixed
 &$8.81^{+0.65}_{-1.28}$ &fixed &167.78/132 & \\ 
 & &[$0.088^{+0.010}_{-0.015}$] & &[$0.321^{+0.019}_{-0.016}$] & & & & &
 & \\ 
E2-m & fixed& $6.019^{+0.067}_{-0.074}$ & $179.6^{+212.9}_{-85.9}$ &
				 $6.524^{+0.035}_{-0.040}$ & $18.6^{+2.9}_{-2.8}$&
 M & $10.61^{+1.13}_{-1.13}$  &fixed &fixed &163.12/132 \\ 
 & &[$0.090^{+0.015}_{-0.014}$] & &[$0.288^{+0.024}_{-0.025}$] & & & & &
 & \\ 
E2-mb & fixed & $6.012^{+0.065}_{-0.083}$ &$195.3^{+288.8}_{-87.5}$ &
				 $6.542^{+0.036}_{-0.030}$ & $21.5^{+3.0}_{-3.1}$ &
M & $6.57^{+1.17}_{-2.14}$ & $5.45^{+1.27}_{-2.20}$
			 &fixed &144.37/131 \\
 & &[$0.089^{+0.014}_{-0.015}$] & &[$0.300^{+0.026}_{-0.020}$] & & & & & \\
E2-mkb &fixed&$6.012^{+0.054}_{-0.080}$ & $193.8^{+280.1}_{-84.2}$
			 &$6.543^{+0.029}_{-0.029}$ &$21.5^{+3.0}_{-3.0}$ & M & $7.16^{+1.71}_{-2.34}$  &$4.93^{+1.48}_{-1.43}$ &fixed &142.54/130  \\ 
 & &[$0.089^{+0.012}_{-0.015}$] & &[$0.301^{+0.021}_{-0.020}$] & &
 K & $<2.49$  & & & \\ 
E2-mkgfb &fixed &$6.016^{+0.060}_{-0.082}$ &$183.4^{+264.3}_{-82.5}$
			 &$6.558^{+0.029}_{-0.031}$ &$22.1^{+3.4}_{-2.8}$ & M & $8.21^{+1.67}_{-2.69}$ &$4.60^{+1.63}_{-1.35}$ &fixed &140.42/129
 \\ 
 & &[$0.090^{+0.013}_{-0.015}$] & &[$0.312^{+0.021}_{-0.021}$] &
 &K& $<3.38$ & & & & \\ 
 & & & & & & FG & $<1.23$ & & & \\ 
E2-(mkgf)b & fixed & $6.007^{+0.037}_{-0.091}$ &
			 $204.2^{+292.6}_{-76.2}$ & $6.519^{+0.027}_{-0.030}$ &
 $21.0^{+3.9}_{-3.6}$ & FGKM & $2.79^{+0.74}_{-0.78}$ &
 $6.22^{+1.35}_{-1.35}$ & fixed & 151.92/129\\ 
 & &[$0.088^{+0.008}_{-0.016}$] & &[$0.285^{+0.018}_{-0.019}$] & & & & &
 & \\ \hline
E2-mb$^{\ddagger}$ & fixed & $6.360^{+0.020}_{-0.021}$ &$33.7^{+3.2}_{-3.2}$ &
				 $\cdots$ & $\cdots$ &
M & $11.77^{+1.11}_{-1.00}$ & $2.50^{+1.43}_{-1.33}$
			 &fixed &214.09/133 \\
 & &[$0.197^{+0.009}_{-0.009}$] & & & & & & & \\
E2-mkgfb$^{\ddagger}$ &fixed &$6.334^{+0.023}_{-0.022}$ & $35.8^{+3.3}_{-3.4}$
			 & $\cdots$ & $\cdots$ &M & 
 $7.01^{+2.64}_{-2.17}$ &$4.34^{+1.52}_{-1.59}$  &fixed &196.99/131
 \\ 
 & &[$0.186^{+0.010}_{-0.009}$] & & & &K& $<2.55$ & & & & \\ 
 & & & & & & FG & $4.29^{+1.31}_{-1.35}$  & & & \\ \hline
\end{tabular}
\begin{minipage}{1.2\textheight}
$^{\dagger}${All parameters are fixed to referred or simulation based
 value without mentioned}\\
$^{\ddagger}${One hot ISM model.}\\
$^{a}${Emission Measure 10$^{-3}$ $\int n_e n_p dl$: in unit of cm$^{-6}$ pc}\\
$^{b}${Spectral type of the stars whose normalization is set
 to be free}\\
$^{c}${Ratio of the normalization of stars of binaries to the
 simulation based value}\\
\end{minipage}
\end{center}
\end{table*}

In the E3  model, an additional unknown component is assumed.
We therefore fixed all the components to the simulated values,  other than the hot ISM.
First, we added a thin thermal plasma to the model because 
the contribution from  an unknown stellar component would be highly possible
(model E3).
The abundance of the additional plasma is fixed to the solar value.
However this model caused  residual features like the E2-b model and 
could not fit the data ($\chi^2$/dof=165.37/131).
The metal abundance of the stellar corona is not well understood and 
so we set the metal abundance to be free (model E3-A).
This model explain the whole spectrum very well
($\chi^2$/dof=121.04/130).
However, the best fit abundance is $\sim0$ and the additional
component is quite close to thermal bremsstrahlung of $kT=$1.1 keV.
Actually we substituted thermal bremsstrahlung for the thin thermal plasma of
E3-A model and found no changes in the fitting results (model E3-B). 
We then set the abundance of the plasma of E3-A model to 0.1 Solar value
(E3-A').
This model also fit the data with $\chi^2$/dof=133.34/131.

Some 1.5 keV thermal components are reported in the spectra of nearby dM
stars (e.g. \citet{vandenBesselaar03}) and this could be the origin of the unknown
component. As shown in table  \ref{tab:StarParameter}, the metallicity of
the coronal spectra of the low luminosity stars is  very low and this is 
consistent with the low metallicity of the unknown component.
Stellar flares are  another possibilitiy for the unknown component.
In the flare period, the high temperature ($> 1$keV) component of the stellar 
corona become brighter than usual.
It is not plausible that this unknown component originates from the hot ISM
because  the metallicity of the unknown component is not high ($<\sim 0.1$) and the temperature and
induced pressure is too high to maintain such plasma.
\begin{table*}
\caption{Fitting results of model E3}
\label{tb:FittingResultsE3}
\footnotesize
 \begin{center}
\begin{tabular}{lccccccccc} \hline \hline
Model& Foreground$^{\dagger}$  & \multicolumn{4}{c}{ISM}
 & \multicolumn{3}{c}{Additional background} &
$\chi^2$/dof\\ \cline{3-6} 
&  and backgrounds &  \multicolumn{2}{c}{cool} &\multicolumn{2}{c}{hot} & &&
 \\ 
&&  $\log T$  (K)& Norm$^{a}$ &$\log T$  (K) & Norm$^{a}$    & $\log T$ (K)  
 & Norm$^{a}$& Abundance  \\ 
 &&[$kT$(keV)] &  &  [$kT$(keV)] & && &  \\ \hline
E3 & fixed& $6.019^{+0.056}_{-0.062}$ & $205.1^{+182.1}_{-87.3}$ &
				 $6.567^{+0.023}_{-0.021}$ & $27.3^{+2.6}_{-2.6}$ &
 $7.544^{+0.104}_{-0.082}$ & $17.4^{+2.1}_{-2.0}$ & 1.0 (fixed) &
 165.37/131 \\
 & &[$0.090^{+0.012}_{-0.012}$] & &[$0.318^{+0.017}_{-0.015}$] &
					 &[$3.017^{+0.816}_{-0.519}$] & & &  \\ 
E3-A &fixed & $6.022^{+0.078}_{-0.072}$ &$171.0^{+190.7}_{-88.4}$
			 &$6.546^{+0.027}_{-0.028}$ &$24.0^{+2.8}_{-3.2}$
					 &$7.109^{+0.105}_{-0.110}$ &$64.3^{+27.3}_{-16.5}$
 &$<0.03$ &121.04/130 \\ 
 & &[$0.091^{+0.018}_{-0.014}$] & &[$0.303^{+0.020}_{-0.019}$] & &
 [$1.109^{+0.303}_{-0.248}$]& & & \\
E3-B &fixed &$6.023^{+0.077}_{-0.071}$ &$169.4^{+185.8}_{-86.9}$
			 &$6.547^{+0.026}_{-0.025}$ &$23.9^{+2.8}_{-2.9}$
					 &$7.106^{+0.108}_{-0.098}$ &$69.6^{+26.2}_{-18.4}$
 & $\cdots$ & 120.94/131\\ 
 & &[$0.091^{+0.018}_{-0.014}$] & &[$0.304^{+0.019}_{-0.017}$] &
					 &[$1.101^{+0.311}_{-0.222}$] & & &  \\ 
E3-A' &fixed &$6.017^{+0.061}_{-0.073}$ &$193.7^{+224.9}_{-87.5}$
			 &$6.548^{+0.026}_{-0.022}$ &$25.9^{+2.7}_{-2.6}$
					 &$7.239^{+0.096}_{-0.077}$ &$38.5^{+4.8}_{-5.8}$
 &0.1 (fixed) &133.34/131 \\ 
 & &[$0.090^{+0.013}_{-0.014}$] & &[$0.305^{+0.019}_{-0.015}$] &
					 &[$1.495^{+0.372}_{-0.243}$] & & & \\ \hline
E3-A'$^{\ddagger}$ &fixed &$6.440^{+0.015}_{-0.015}$ & $37.3^{+2.3}_{-2.3}$
			 & $\cdots$ & $\cdots$
					 &$7.151^{+0.038}_{-0.039}$ &$49.8^{+4.7}_{-4.7}$
 &0.1 &231.45/133 \\ 
 & &[$0.237^{+0.008}_{-0.008}$] & & &
					 &[$1.221^{+0.112}_{-0.104}$]& & &  \\ \hline
\end{tabular}
\end{center}
$^{\dagger}${All parameters of LHB+SWCX, Loop I, stars, and CXBs are fixed to referred 
or simulation based  values }\\
$^{\ddagger}${One hot ISM model.}\\
$^{a}${Emission Measure 10$^{-3}$ $\int n_e n_p dl$: in unit of cm$^{-6}$ pc}\\

\end{table*}

\begin{figure}
\begin{center}
\FigureFile(0.45\textwidth, 80mm){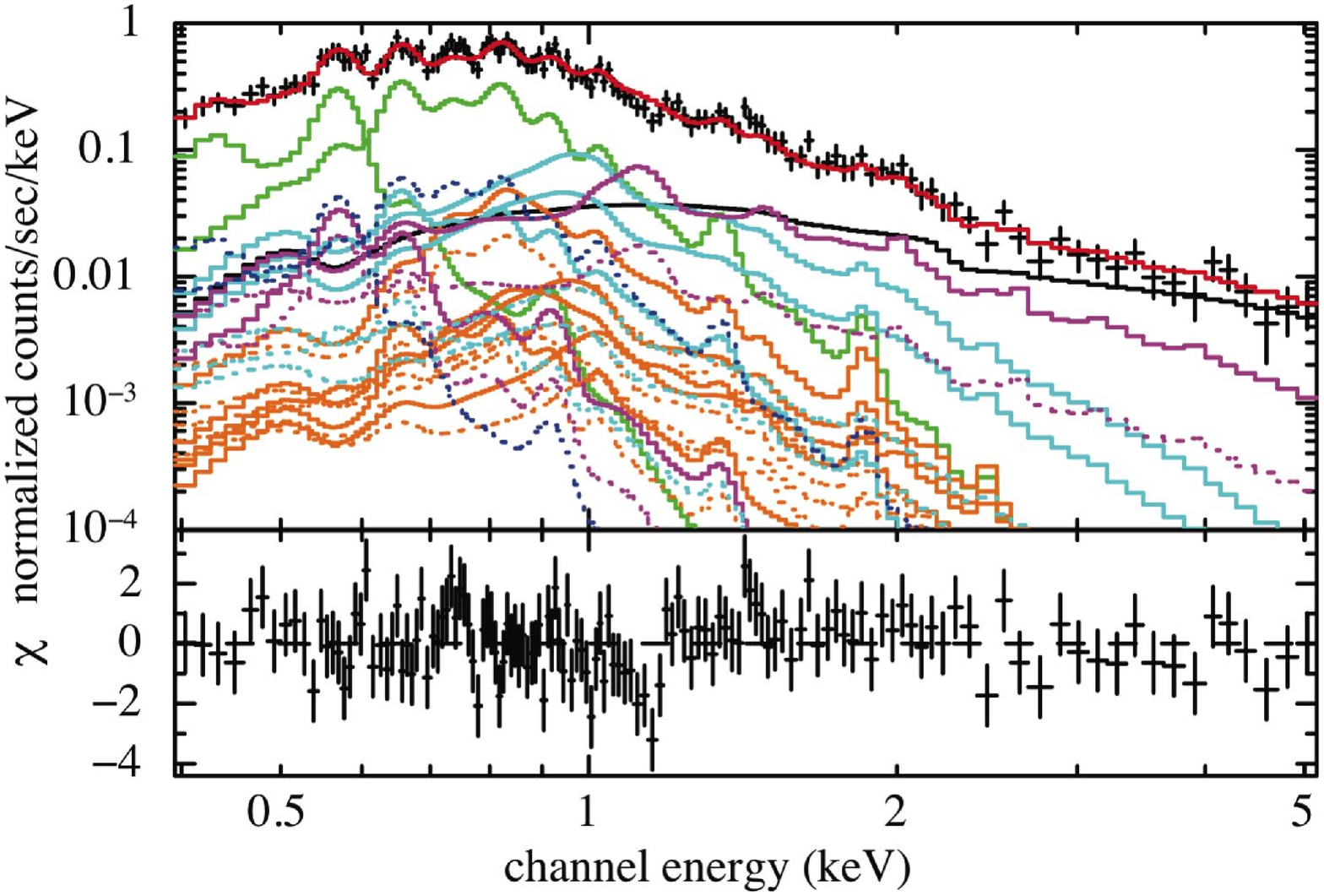}
\FigureFile(0.45\textwidth, 80mm){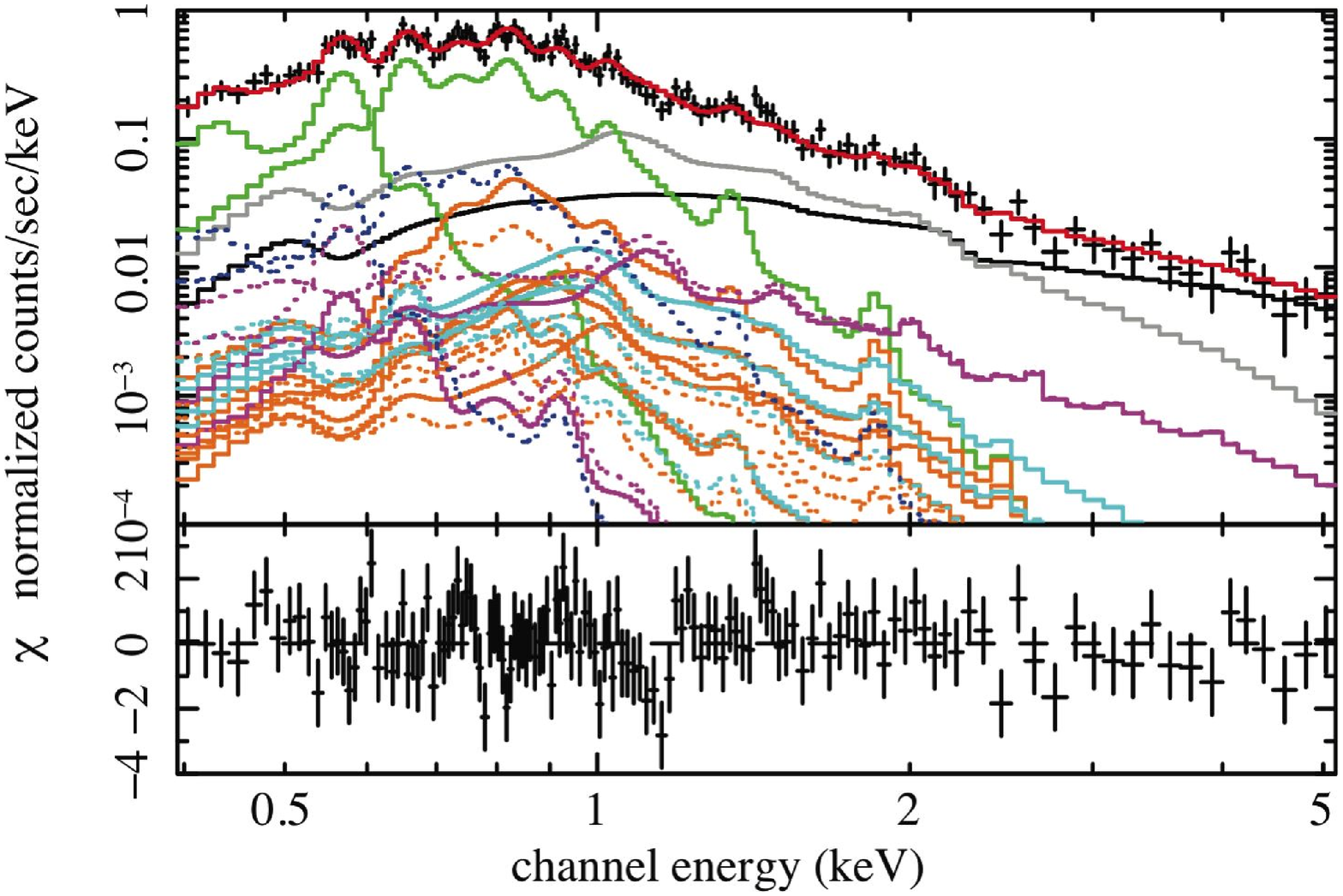}

\caption{Best-fitted model and spectra by model E2-mb and E3-A'
The dotted lines indicate foreground
 (unabsorbed) components and the solid lines indicate background (fully
 absorbed) components. hot ISM (green), CXB (black),  SWCX+LHB, 
 and Loop I (blue), stars except for M stars (orange), M stars (light blue),
and RS CVns (magenta).Additional background continuum in E3 model is indicated by 
gray line.}
\end{center}
\label{fig:E-bestmodel}
\end{figure}

From model E2 and model E3, we see that $\sim 5$ times normalization of
stellar components or
another thermal bremsstrahlung or a low abundance thin thermal plasma is needed to
explain the whole spectrum.
As shown in table \ref{tb:FittingResultsE2} and
\ref{tb:FittingResultsE3}, there is at most a 10\% difference in hot
ISM parameters between acceptable models.
Our model assumes two extreme possibilities and it is reasonable to 
consider that the star contribution would change the hot ISM parameters
at most by 10\% in any case.
Thus we use model E2-mb and E3-A' to represent the  emission model 
in the further analysis.

We evaluated the line intensities corresponding to the hot ISM component 
with model E2-mb and E3-A'  as summarized in table \ref{tb:SurfacebrightnessE2}.
The intensity and ratio of the O lines are  little affected  by the stellar or additional background 
 models, but the 
contribution of the Ne lines  changes by the  assumption of stellar components, 
especially of the high temperature stars.

\begin{table}
\caption{Surface brightness of the O and Ne lines.}
 \label{tb:SurfacebrightnessE2}
\begin{center}
\begin{tabular}{ccccc} \hline \hline
Model & O\emissiontype{VII} & O\emissiontype{VIII} & Ne\emissiontype{IX} &  Ne\emissiontype{X}\\ 
& (LU) &(LU) &(LU) & (LU)\\\hline
E2-mb & $14.72^{+2.74}_{-1.47}$ & $9.80^{+-0.22}_{-3.15}$ &
			 $0.61^{+1.29}_{-0.47}$ & $<0.19$ \\
E3-A' & $15.80^{+2.73}_{-1.51}$ & $10.75^{+1.48}_{-1.32}$ &
			 $3.09^{+1.05}_{-0.54}$ & $2.17^{+1.05}_{-1.24}$ \\ \hline
\end{tabular}
\end{center}
\end{table}

\subsection{Combined analysis}

Two thermal components are at least required to describe the 
emission spectrum.
Assuming the geometry of the two temperature plasmas and the absorption
toward the  target, there are  two possibilities for combined analysis.
One is that only one plasma contributes  the absorption, and the other is 
that  both plasma contribute to the absorption.
Thus, we consider these two cases in this  section.

\subsubsection{Uniform  model with one absorbing plasma }
\begin{figure}[h]
\begin{center}
\FigureFile(80mm, 60mm){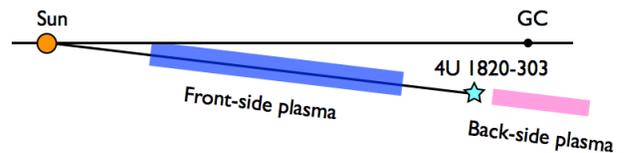}

\end{center}
\caption{Schematic view of the model C1.}
\label{fig:IllustC1}
\end{figure}

We assumed first as a combined model that only one plasma contributes to  the
absorption (model C1).
The geometry is as follows; one uniform plasma (front-side plasma) exists in front of 4U
1820$-$303 and another uniform plasma (back-side plasma) exists in back of 4U
1820$-$303, as illustrated in figure \ref{fig:IllustC1}.

We constructed two isothermal plasma models with a uniform density, and 
a length $L$ along the sight of line.
We put an upper limit of 7.6 kpc on the length of the front-side plasma, 
to maintain consistency with the geometry.
The velocity dispersions of the plasmas are linked together 
 because this value
was mainly determined using the ratio of   O\emissiontype{VII} K$_{\alpha}$ to O\emissiontype{VII}
K$_{\beta}$ in the absorption spectrum. 
We tried four sets of models, C1-1 and C1-2 is  the combination of E2-mb and A4 model, and 
Ne/O abundance ratio is  set to be free in C1-2. 
In C1-3 and C1-4 model, we adopt the E3-A' model for the emission model, and Ne/O ratio 
is set to be free in C1-4.

The fitting results are shown in table \ref{tb:FittingResultsC1}.
The temperature of the front-side plasma (T$\sim1.7\times10^{6}$ K) is determined mainly by the
absorption spectra and is consistent with the temperature
obtained by the absorption analysis (T$\sim1.8\times10^{6}$ K).
The temperature of the front- and back-side plasma ($1.7\times10^6$ K and
$3.9\times10^6$ K) are both higher
than the those determined only by emission analysis ($1.1\times10^6$ K and
$3.6\times10^6$ K).
Plasma of $1.7\times10^6$ K could emit O\emissiontype{VII} and O\emissiontype{VIII}
lines three times and thirty times effectively than a plasma of
$1.1\times10^6$ K.
Though this makes the emission measure of the front-side plasma smaller
to maintain the intensity of the O\emissiontype{VII} lines,
O\emissiontype{VIII} lines are produced more effectively.
Thus to suppress the O\emissiontype{VIII} intensity, the temperature of the
back-side plasma becomes higher.
Residuals caused by this adaptation could be compensated by the 
background components and this model also can reproduce the O, Fe and Ne emission lines.
However in the energy range lower than 0.5 keV, there are  residuals
caused by the deficit of N lines, because the lower temperature plasma
($1.7\times10^6$ K) is too hot to emit N lines effectively.

\begin{table*}
 \caption{Fitting results of C1 model}
\label{tb:FittingResultsC1}
\scriptsize
 \begin{center}
\begin{tabular}{llcccccccccccc} \hline \hline
\multicolumn{2}{c}{Model}& \multicolumn{5}{c}{Front-Side plasma} & \multicolumn{2}{c}{Back-Side plasma}   & Stars
 & Binaries  & \multicolumn{2}{c}{Additional} &$\chi^2$/dof \\ \cline{3-7}
 & &$N_{\rm H_{Hot}}$ & Length &  $\log T$ & Ne/O &$v_b$ & $\log T$ & Norm$^{a}$
 &Ratio$^{b}$&Ratio$^{c}$& $\log T$ & Norm\\ 
  & & (cm$^{-2}$) &(kpc) & (K) & &(km s$^{-1}$) &  (K)  & & & &(K) \\ \hline
C1-1 & E2-mb & $19.71^{+0.09}_{-0.37}$ &$7.60^{+0.00}_{-5.61}$&
				$6.22^{+0.05}_{-0.08}$  &$\cdots$ &$\cdots$ & $6.583^{+0.055}_{-0.049}$
&  $15.5^{+5.1}_{-4.9}$  & $7.00^{+1.90}_{-2.83}$ & $4.66^{+1.56}_{-1.51}$ 
&$\cdots$&$\cdots$ & 819.82/704\\
& A4 &$\uparrow$ &$\cdots$ &$\uparrow$ & $\cdots$&  $109^{+140}_{-29}$\\
C1-2 & E2-mb  &$19.60^{+0.18}_{-0.22}$ & $4.67^{+2.93}_{-2.91}$ &$6.22^{+0.05}_{-0.06}$
				& $1.4^{+1.2}_{-0.8}$  
				  &$\cdots$ 
					 &$6.588^{+0.047}_{-0.051}$&$15.3^{+5.2}_{-3.9}$  &
 $6.67^{+1.98}_{-1.59}$ & $4.86^{+1.48}_{-1.53}$   &$\cdots$&$\cdots$&819.05/703\\ 
 &A4 &$\uparrow$ &$\cdots$ &$\uparrow$ &$\uparrow$& $126^{+164}_{-45}$
						 &$\cdots$  &$\cdots$  & & & &  \\ 
C1-3 & E3-A' & $19.68^{+0.16}_{-0.11}$ & $5.88^{+1.72}_{-3.66}$
			 &$6.20^{+0.05}_{-0.04}$ &$\cdots$ &$\cdots$ &$6.574^{+0.041}_{-0.029}$
						 &$22.0^{+3.0}_{-5.2}$  &$\cdots$ &$\cdots$
 &$7.254^{+0.085}_{-0.086}$ & $37.7^{+5.2}_{-5.3}$ &810.11/704 \\ 
 & A4 &$\uparrow$ &$\cdots$ &$\uparrow$ & $\cdots$&$111^{+84}_{-34}$ & &
 & & & & & \\ 
C1-4 &E3-A' &$19.53^{+0.21}_{-0.19}$  & $3.08^{+4.52}_{-1.81}$&
				$6.23^{+0.05}_{-0.05}$&$2.3^{+1.4}_{-1.1}$ &$\cdots$ &
							 $6.592^{+0.040}_{-0.040}$&$18.9^{+5.1}_{-4.0}$
 &$\cdots$ &$\cdots$ &$7.265^{+0.089}_{-0.073}$ &$37.1^{+5.0}_{-5.5}$ &805.94/703\\ 
 &A4 &$\uparrow$ &$\cdots$ &$\uparrow$ &$\uparrow$&$139^{+195}_{-55}$ & & & & & & & \\ \hline
\end{tabular}
\end{center}
Note: All the parameters not written the table is fixed to the
 simulation based values.\\
$^{a}${Emission Measure 10$^{-3}$ $\int n_e n_p dl$: in unit of cm$^{-6}$ pc}\\
$^{b}${Normalization of background M type star is set
 to be free}\\
$^{c}${Ratio of the normalization of the background stars or binaries to the
 simulation based value}\\
\end{table*}

We summarized the physical properties of the plasma in table 
\ref{tb:PhysicalPropertiesC1}.
We assume 1, 2 and 10 kpc for  the length for the back-side plasma to
estimate its density and pressure, because the back-side plasma 
contributes to the emission, and we cannot determine the length and density separately.
There could be two schemes.
Assuming the back-side plasma confined in the Galactic bulge region,
its length is at most 2 kpc and this leads to dense and high pressure plasma.
Thus, this leads to a picture of that a hotter dense plasma existing around the Galactic center region and 
a warm thin plasma covering  the disk.
The other  is that assuming pressure equilibrium between front-side plasma and
back-side plasma, the length of the back-side plasma becomes $\sim$ 8kpc,
This means that a hotter plasma of large depth  exists over the warm thin disk, 
because 4U 1820$-$303 lies 1 kpc below the Galactic disk.

\begin{table*}
\caption{Physical properties obtained by the model C1-4}
\label{tb:PhysicalPropertiesC1}
\begin{center}
\begin{tabular}{llccccc} \hline \hline
Model & Component &  Length & Density & Temperature& Pressure \\
&& (kpc) & ($10^{-3}$cm$^{-3}$) & ($10^6$ K)&($10^{3}$cm$^{-3}$ K) \\\hline
C1-4& Front-Side plasma &$3.08^{+4.52}_{-1.81}$ & $3.6^{+10.4}_{-2.6}$ &$1.7^{+0.2}_{-0.2}$ & 
			 $6.0^{+20.6}_{-4.6}$\\
 & Back-Side plasma & 1  & $4.3^{+0.6}_{-0.4}$ &$3.9^{+0.4}_{-0.3}$ & $16.8^{+4.2}_{-2.9}$ \\
	&	 & 2  & $3.1^{+0.6}_{-0.4}$ &$\uparrow$
				 &$12.1^{+3.7}_{-2.5}$\\
 & & 10 & $1.4^{+0.2}_{-0.2}$  & $\uparrow$ & $5.1^{+1.3}_{-1.2}$\\ \hline
\end{tabular}
\end{center}
\end{table*}

\subsubsection{Uniform plasma model of two absorbing  component}

\begin{figure}[h]
\begin{center}
\FigureFile(80mm, 60mm){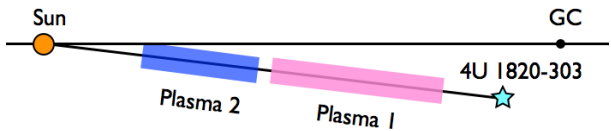}
\end{center}
\caption{Schematic view of the model C2.}
\label{fig:IllustC2}
\end{figure}

For the second  step, we assumed that two   plasma components 
which exist in front of  4U 1820$-$303 contribute  both 
the absorption and emission  (model C2)  as illustrated in figure \ref{fig:IllustC2}.
Note that this assumes that there is no emission from the bulge behind 4U 1820$-$303.
It is not determined whether  the two plasma components are separate, 
or mixed together with some  filling factors,
but we again set an upper limit for the length to be 7.6 kpc and a common velocity dispersion.

As with the  previous model C1, 
we constructed two models with model sets of E2-mb + A4 (C2-1, C2-2) and
E3-A' + A4 (C2-3, C2-4).
The fitting results are shown in table \ref{tb:FittingResultsC2}.
These values can be understood as follows; 
First, two  plasma components of temperatures  $\sim3.5\times10^{6}$ K
and $\sim1.0\times10^{6}$ K
are required to reproduce  the emission spectra.
The ionization fraction and emissivity of each plasma is determined by their temperature,  then
the  column density and the emission
measure to reproduce the absorption  spectrum are obtained.
We confirm this flow and found that the emission measures obtained here 
are about half of those obtained in the emission analysis.
This is caused by the slight temperature decrease of hotter plasma
induced by this  combined analysis.

With the model C2 results and the assumption of the  
 Ne abundance of   solar value,
 and the temperature dependence of the  ionization fraction, 
the Ne\emissiontype{X} column density is
 at most $1\times10^{15}$ cm$^{-2}$.
The upper limit of the column density of Ne\emissiontype{X}
from that of EW in the absorption spectrum  in table \ref{tab:abs_ew}  is 
$\log N_{{\rm Ne X}} = 15.4$, and is consistent with this upper limit.

\begin{table*}
\caption{Fitting results with C2 model}
\label{tb:FittingResultsC2}
\scriptsize
\begin{center}
\begin{tabular}{llcccccccccccccc} \hline \hline
\multicolumn{2}{c}{Model} & \multicolumn{5}{c}{Uniform hot plasma 1} &  \multicolumn{3}{c}{Uniform hot plasma 2} & Stars
 & Binaries  &\multicolumn{2}{c}{Additional}  &$\chi^2$/dof \\ \cline{3-7}
 & &$N_{\rm H_{Hot}}$ & Length &  $\log T$ & Ne/O &$v_b$ &$N_{\rm H_{Hot}}$ &
 Length &  $\log T$   &Ratio$^{a}$&Ratio$^{b}$&$\log T$ & Norm$^{c}$\\
 & & (cm$^{-2}$) &(kpc) & (K) & &(km s$^{-1}$) & (cm$^{-2}$) &(kpc) &(K)  & & &(K) & & \\ \hline
C2-1 & E2-mb & $19.55^{+0.04}_{-0.24}$ &$7.6^{+0.0}_{-5.2}$ &
				 $6.48^{+0.04}_{-0.04}$&$\cdots$  &$\cdots$  &$19.26^{+0.33}_{-0.32}$&$0.17^{+0.66}_{-0.15}$
						 & $5.99^{+0.07}_{-0.15}$ &
 $7.58^{+1.92}_{-1.81}$  & $4.73^{+1.47}_{-1.46}$&$\cdots$ &$\cdots$ & 805.65/703\\
& A4 &$\uparrow$ &$\cdots$ &$\uparrow$ & $\cdots$  &
						 $139^{+286}_{-44}$ &$\uparrow$ & $\cdots$ &$\uparrow$  \\ 
C2-2 &E2-mb &$19.55^{+0.04}_{-0.22}$ & $7.6^{+0.0}_{-5.1}$
			 &$6.48^{+0.04}_{-0.04}$ &$0.9^{+0.4}_{-0.4}$& $\cdots$ 
						 
&$19.26^{+0.34}_{-0.31}$ & $0.18^{+0.67}_{-0.16}$ &$5.99^{+0.07}_{-0.15}$
 & $7.85^{+2.01}_{-2.09}$ & $4.56^{+1.61}_{-1.62}$
 &$\cdots$&$\cdots$ & 805.49/702  \\ 
 &A4 &$\uparrow$ &$\cdots$ &$\uparrow$ & $\cdots$  &$140^{+282}_{-44}$
						  &$\uparrow$ & $\cdots$ &$\uparrow$  \\ 
C2-3 &E3-A' &$19.57^{+0.03}_{-0.22}$ &$7.5^{+0.1}_{-4.9}$ &
				 $6.51^{+0.02}_{-0.03}$ &$\cdots$  &$\cdots$  & $19.22^{+0.37}_{-0.25}$&
						 $0.2^{+0.7}_{-0.1}$ &$6.01^{+0.07}_{-0.10}$ &
$\cdots$ &$\cdots$ &$7.247^{+0.069}_{-0.094}$ &$40.6^{+5.0}_{-4.7}$ &800.67/703  \\
 &A4 &$\uparrow$ &$\cdots$ &$\uparrow$ & $\cdots$ &$156^{+269}_{-64}$ &$\uparrow$ & $\cdots$ &$\uparrow$ 
 & & \\
C2-4 &E3-A' &$19.52^{+0.16}_{-0.21}$ & $5.9^{+1.7}_{-3.7}$
			 &$6.50^{+0.03}_{-0.04}$&$1.4^{+0.4}_{-0.3}$ & $\cdots$ 
						 &$19.19^{+0.33}_{-0.25}$&$0.13^{+0.56}_{-0.11}$&
 $6.00^{+0.07}_{-0.11}$&$\cdots$ &$\cdots$
 &$7.270^{+0.082}_{-0.069}$
 &$39.0^{+5.2}_{-5.3}$ &796.36/702 & \\ 
 &A4 &$\uparrow$ &$\cdots$ &$\uparrow$ & $\uparrow$   &$169^{+273}_{-76}$ &$\uparrow$ & $\cdots$ &$\uparrow$ &  \\ \hline
\end{tabular}
\end{center}
Note: All the parameters not written the table is fixed to the
 simulation based values.\\
$^{a}${Normalization of background M type star is set
 to be free}\\
$^{b}${Ratio of the normalization of the background stars or binaries to the
 simulation based value}\\
$^{c}${Emission Measure 10$^{-3}$ $\int n_e n_p dl$: in unit of cm$^{-6}$ pc}\\

\end{table*}

We summarized the induced physical properties of the plasma 
in table \ref{tb:PhysicalPropertiesC2}.
The length of the hotter plasma is almost on the upper limit.
While the cooler component has very short length $<0.7$ kpc and 
the pressure is higher than that of the hotter by a factor of $\sim$ 7.
This assumption gives a scheme of a thin warm  and a thick hot disk halo model.

\begin{table*}
\caption{Physical properties obtained by model C2-4}
\label{tb:PhysicalPropertiesC2}
\begin{center}
\begin{tabular}{llccccc} \hline \hline
Model & Component &  Length & Density &Temperature &  Pressure \\
&& (kpc) & ($10^{-3}$cm$^{-3}$) & ($10^6$ K) & ($10^{3}$cm$^{-3}$ K) \\\hline
C2-4& Plasma 1 & $5.9^{+1.7}_{-3.7}$ &$1.8^{+5.2}_{-0.9}$ &$3.2^{+0.2}_{-0.3}$ & $5.7^{+18.1}_{-3.2}$ \\
& Plasma 2 & $0.13^{+0.56}_{-0.11}$  & $38.6^{+497.3}_{-34.5}$ &$1.0^{+0.2}_{-0.2}$ & 
			$38.6^{+591.0}_{-35.4}$ \\ \hline
\end{tabular}
\end{center}
\end{table*}

\subsection{Uncertainty due to the model systematics}
In the absorption analysis, the energy resolution of the detectors ($\sim0.05{\rm \AA}$) corresponds
to $ v_b \sim 400{\rm km s^{-1}}$ and is not enough to determine the
$v_b$ of the plasma only with the line shapes. 
The lower limit of  $v_b$ can be determined from thermal limits.
Thus, to determine the velocity dispersion, we used the ratio
of the absorption depth of O\emissiontype{VII} K$_{\alpha}$ and  O\emissiontype{VII}
K$_{\beta}$ instead and assumed  that the O\emissiontype{VIII}, 
Ne\emissiontype{IX} and Ne\emissiontype{X} originate from the same plasma, and 
linked the velocity dispersion of all lines.
As  $v_b$ and column density are
coupled and 
if the plasma has a  temperature gradient and local structure,
 this assumption causes systematic errors, but it effects little in the 
 temperature of the ISM as shown by \citet{hagihara10}.

In table  \ref{tb:SystematicModel} we summarize the systematic errors of
the background and foreground models and their effect on the best fit values of the hot ISM parameters.
The obtained hot ISM parameters are all within 90 \% statistical error
and cause no change to our results and conclusions.

\begin{table*}
\caption{Estimated systematic errors of the models and their effect on
 the best fit values of the hot ISM parameters (all the parameters are
 within statistical error).}
\label{tb:SystematicModel}
\begin{center}
\begin{tabular}{lccccc} \hline \hline
 Component &  Systematic  & \multicolumn{4}{c}{Affection to the best fit values
 of the Hot ISM} \\ 
 &Error & \multicolumn{2}{c}{Cool ($T<1.3 \times 10^6$ K)} &
 \multicolumn{2}{c}{Hot ($T>2.5 \times 10^6$ K)}  \\ 
 & & Temperature & Normalization & Temperature & Normalization \\\hline
Foreground stars & $+20\%$ & $<1\%(+)$ &  $-3\%$ & $<1\%(-)$ &  $<1\%(-)$  \\
 & $-20\%$ & $<1\%(-)$ &  $+3\%$ & $<1\%(+)$ &  $<1\%(+)$  \\ \hline
Background stars & $+50\%$ & $<1\%(-)$ &  $+2\%$ & $<1\%(-)$ &  $-2\%$ \\
 & $-50\%$ & $<1\%(+)$ &  $-2\%$ & $<1\%(+)$ &  $+2\%$  \\ \hline
CXB & $+30\%$ & $<1\%(+)$ &  $\pm3\%$ & $<1\%(-)$ &  $\pm1\%$ \\
 & $-30\%$ &  $<1\%(-)$ &  $\pm3\%$ & $<1\%(+)$ &  $\pm1\%$\\ \hline
LHB+SWCX & $+75\%$ &  $<1\%(+)$ &  $-15\%$ & $<1\%(+)$ &  $<1\%(\pm)$ \\
 & $-100\%$ &  $<1\%(-)$ &  $+18\%$ & $<1\%(+)$ &  $<1\%(\pm)$\\ \hline
Loop I & $+50\%$ & $<1\%(+)$ &  $-3\%$ & $<1\%(-)$ &  $-10\%$ \\
 & $-50\%$ & $<1\%(-)$ &  $+4\%$ & $<1\%(+)$ &  $+10\%$\\ \hline
\end{tabular}
\end{center}
\end{table*}

\begin{figure}[h!]
\begin{center}
\FigureFile(80mm, 50mm){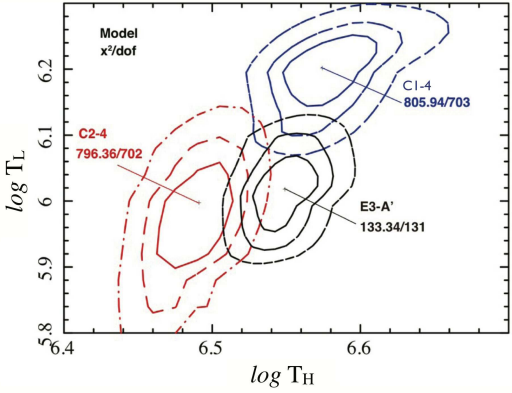}

\end{center}
\caption{68\%, 90\%, and 99\% confidence contours of the temperature
 of hot ISM for two parameters.
 Black, red and blue lines indicate the results of E3-A' model, C1-4 model and
 C2-4 model respectively. The best fitted $\chi^2$/dof values are indicated. }
\label{fig:ModelSystematic}
\end{figure}

In addition to the systematic errors mentioned above,
we have to consider systematics caused by the combined analysis itself.
Figure  \ref{fig:ModelSystematic} shows contours of the temperatures of the
hot ISM derived from emission analysis (E3-A') and combined
analysis (C1-4, C2-4).
As shown in figure  \ref{fig:ModelSystematic}, contours only overlap partially and those from
combined analysis are shifted to each direction.
The best fitted $\chi^2$/dof value of C1-4 model is worse than those of
the other models.
These shifts would be caused by modeling a multi-temperature plasma with
two temperature plasmas. 
To check the contribution of the uncertainty of the star and
unknown component, we fixed the parameters of the unknown component to
the best fit value of E3-A' model (table \ref{tb:FittingResultsE3}) and
performed a combined analysis with the C1-4 and C2-4 models.
The results are consistent with  those of C1-4 and C2-4 model.
These two models (C1-4, C2-4) are representative of the other combined analysis
models and  we concluded the systematics originate in the combined analysis. 
Using these contours, the temperature range of the two plasmas are
determined as $2.8-4.3 \times 10^6$ K and 
$0.68-1.9 \times 10^6$ K.

We also checked the self-absorption effect of the emission lines with the obtained column density and velocity 
dispersion of ions. When the column density O\emissiontype{VII}  is $10^{16}$cm$^{-2}$ and the minimum  velocity 
dispersion $v_{b} = 200 $ km s$^{-1}$, the opacity $\tau$ at the center of the resonance line is $\sim$1.1, which is 
 evaluated in the  same way as  \citet{futamoto04}.  
The reduction due to the self-absorption for the resonance line is about 30\% after integrating over the 
emission line profile. Note that we measured all resonance, inter-combination,  and forbidden lines 
as O\emissiontype{VII} line, with the energy resolution of the Suzaku XIS. As the oscillator strength of the forbidden line 
is very small, the apparent reduction is diluted to be half as small as 15\%. Also we neglected the effect of scattered--in 
photons from outside the line of sight.  The $\tau$ at the  O\emissiontype{VIII} line center is less than the unity with 
$b = 200 $ km s$^{-1}$ and $N_{\rm O\emissiontype{VIII}} = 10^{16}$ cm$^{-2}$, and the total reduction is less than 20\% 
without considering the O\emissiontype{VII} K$\beta$ line.  These effects will systematically reduce the emission 
line intensity, but we neglect the effect in this paper because the correction is as much as the systematic and 
statistical errors. 
When we obtain the better energy resolution for the diffuse emission, 
we will able to evaluate by the ratio between the resonance and forbidden lines.

\section{Discussion}

In this paper, we analyzed  emission and absorption spectra toward the Galactic bulge region, 
and found  that there are at least two models
to explain the emission and absorption data in this direction.
We summarized the results in table \ref{tb:BulgeResults}.
Only the emission
measures are obtained  for the back-side component, because there are no absorption data.
As we simplified the models as much as possible, we will discuss  the hot plasma 
within our Galaxy by possible extension from current simple models and by comparing the
previous results at high latitude. 

\subsection{Extension of the current simplest geometry}
\begin{figure}[h]
\begin{center}
\FigureFile(80mm, 60mm){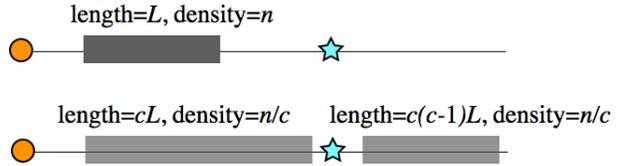}
\end{center}
\caption{Schematic view of the extended plasma model beyond 4U1820$-$303. }
\label{fig:extend}
\end{figure}

First,  we took notice of the length of the front component.
It is not likely that such a fully filled component is confined only in a sight line, 
and this implies that these components fill and prevail throughout the disk.
Thus,  we consider that such long extended plasma is a part of the hot ISM disk.
Except for Plasma 2 in C2 model, the error range includes the  upper limit determined by the geometry, 
which is the distance of 7.6 kpc toward  4U1820$-$303. It is not reasonable that the 
boundary of the two  plasmas coincide at   4U1820$-$303.
It is useful to consider what happens when a plasma of the same temperature 
exists beyond 7.6 kpc as in  figure \ref{fig:extend}.
Additional plasma at the backside of 4U 1820$-$303, 
which does  not contribute to the absorption, makes the
emission measure  of the original plasma decrease to maintain the emission intensity,
while the column density  is constant.
The length becomes longer and the density becomes smaller assuming
additional distant  plasma.
If the density of absorbing plasma is reduced by a  factor of $1/c$, 
and the length is increased by the same factor $c$, the column density 
is kept as the same and the emission measure is reduced by 
 $1/c$. Thus if  there is  emitting plasma behind the target 
with density $n/c$ and length of  $c(c-1)$ times the original one, the emission measure is 
recovered. It is easily written  with the density $n$, the length $L$ and a 
reduction factor $c$, 
\begin{equation}
CD = n/c \times cL=nL \label{eq:extCD}
\end{equation}
\begin{equation}
EM=(n/c)^{2}\times cL + (n/c)^{2}\times c(c-1)L = n^{2}L \label{eq:extEM}
\end{equation}

\begin{figure}[h]
\begin{center}
\FigureFile(0.45\textwidth, 60mm){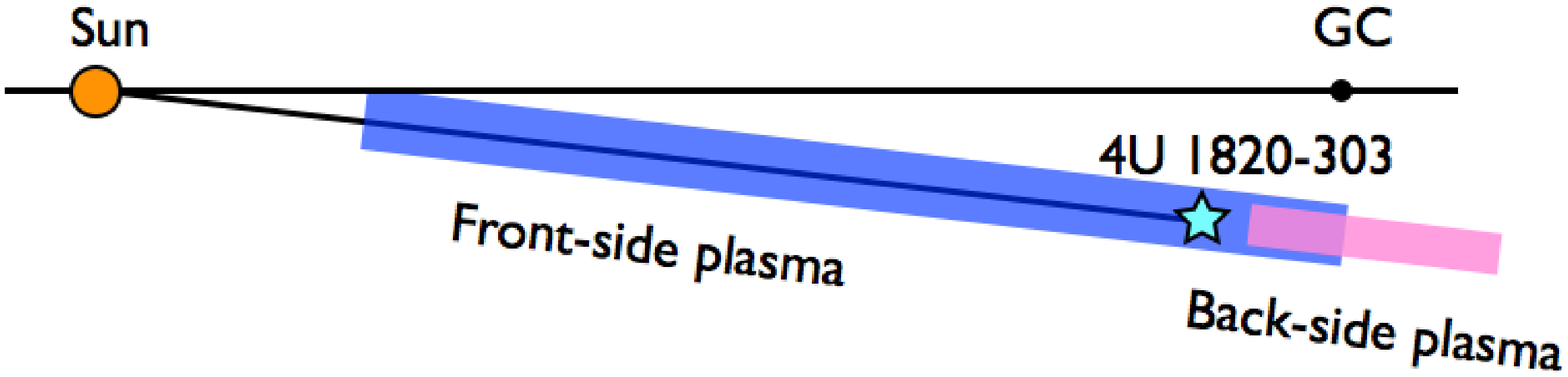}
\FigureFile(0.45\textwidth, 60mm){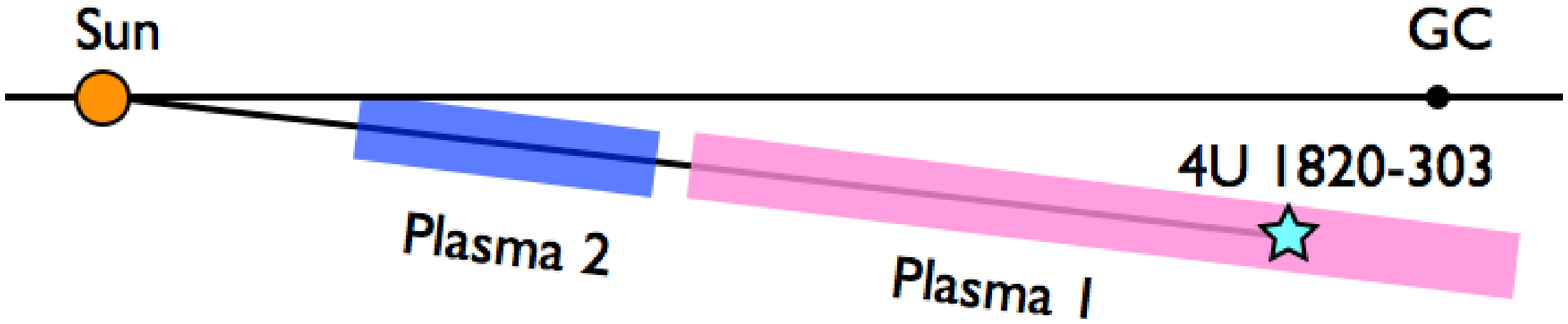}

\end{center}
\caption{Elongated models correspond  to  case C1 and C2 (C1-B and C2-B) }
\label{fig:CaseB}
\end{figure}

Both C1 and C2 models can be  extrapolated like figure \ref{fig:CaseB} as 
case C1-B and C2-B.

\subsection{Comparison with the exponential disk  observed at the high latitude }
The C1 model implies  a thin warm disk with a length of   $3.1$ kpc 
in the front and a hot plasma at the back, toward the bulge region.
The hot plasma is possible to associate with the bulge, or to be  
a hot thick disk above the warm disk. 
We will compare the properties of the warm disk with 
 the exponential disk model obtained by 
the combined analysis with absorption to  extragalactic objects; PKS2155$-$304 
\citep{hagihara10} and LMC X-3 \citep{yao09}.
In the extreme case, we can assume that 
the warm  plasma has an extent of 7.6 kpc which corresponds to $c=7.6/3.1=2.5$ in
the above equations \ref{eq:extCD} and \ref{eq:extEM}. 
The total length and the total column density are calculated as $(c L +
c(c-1) L)=c^2L=19.4$ kpc
and $nL+(c-1)nL=cnL=10^{19.93}$ cm$^{-2}$.
The thickness of the disk is 19.4$\times{\rm sin}(b_{\rm 4U})=2.7$
kpc, where $b_{\rm 4U}=-7.9^{\circ}$ is the Galactic latitude of 4U 1820$-$303.
This is comparable to that of the exponential disk.
The column density is also comparable to that found for  PKS2155$-$304
considering   the difference of the Galactic latitude, as 
$10^{19.93}\times{\rm sin}(b_{\rm 4U})/{\rm sin}(b_{\rm
PKS})=10^{19.17}$ cm$^{-2}$, 
where $b_{\rm PKS} = -52.2 ^{\circ}$ is  the Galactic latitude of 
PKS 2155$-$304. This value is consistent with the column density 
for the PKS 2155$-$304 direction, $\log N_{\rm H}=10.10 ^{+0.08}_{-0.07}$  obtained 
by an exponential disk model \citep{hagihara10}.
Thus the front plasma  of C1 model  corresponds to
the exponential disk.

To confirm this similarity, we applied an exponential disk model with parameters,  
which represent well both  the emission and absorption spectra of PKS 2155$-$304
\citep{hagihara10}.  The exponential model well matched the absorption spectrum, 
but failed for the emission spectrum due to the residuals  below 1 keV. 
If we add components which only contribute to the emission like C1 model, 
the fit returns a $\chi^{2}=809.8/706$ with two thermal component with $log T=
6.597^{+0.046}_{-0.044}$ and $6.023^{+0. 251}_{-0.183}$ with a stellar contribution 
the same as in  E3-A'.

The hot plasma at the backside was not observed at high latitude 
with LMC X-3 and PKS2155$-$304 \citep{yao09, hagihara10}, 
but was detected  in observations toward other bulge regions \citep{almy00}. 
The temperature and estimated electron density assuming the size of 10 kpc 
is  consistent with those  by the RASS image model by \citet{snowden97}.
These support the idea that the back-side plasma is a hot plasma 
associated with the bulge region. 
The location, size, and pressure are, however,   
not determined by current observations, and the relation between the disk and the bulge is hard to be considered.

The C2 model implies a  thin warm and thick hot disk halo model.
Such a thin disk has not been observed by previous studies, and its pressure 
is as high as $4\times10^{4}$cm$^{-3}$K, and almost twice of the typical 
value at the midplane estimated by  \citet{Cox05} with thermal and non-thermal
(Cosmic rays and magnetic field) components. 
The pressure of the warm plasma is also  not balanced by a factor of 6. 
In addition, this simple C2 model assumes that the 
hot ISM is spatially limited in front of 4U1820-303 and no plasma in the bulge region.
If  the hotter plasma can be extended beyond 4U1820$-$303 like C2-B in figure \ref{fig:CaseB},
the pressure will decrease in proportional to the  density.
We thus need some mechanism to confine the warm $T\sim10^{6}$ K plasma close to the 
Galactic disk.

\begin{table*}
\begin{center}
\caption{ Results of the analysis of the galactic bulge region.
}
\label{tb:BulgeResults}
\begin{tabular}{llccccccc}\hline\hline
Model & Component &  \multicolumn{3}{c}{Front-Side} & \multicolumn{2}{c}{Back-Side}   & \\
 & & $T$  & $\log N_{\rm H_{Hot}}$ &Length & $T$ & EM  \\ 
 & &($10^{6}$K) &(${\rm cm^{-2}}$) &(kpc) &($10^{6}$K) &($10^{-3} {\rm cm^{-6} pc}$) & \\ \hline
C1 & Front-Side Plasma & 1.7$\pm$0.2  &19.53$\pm0.2$ & 3.1$^{+4.5}_{-1.8}$ &$\cdots$&$\cdots$& 
				 \\
 &Back-Side Plasma	&  $\cdots$&$\cdots$&$\cdots$& 3.9$^{+0.4}_{-0.3} $& 18.9$^{+5.1}_{-4.0}$	\\
C2  & Plasma 1 & 3.2$^{+0.2}_{-0.3}$ & 19.52$^{+0.16}_{-0.21}$ & 5.9$^{+1.7}_{-3.7}$ &
				  $\cdots$&$\cdots$& \\
 &Plasma 2 & 1.0$\pm0.2$ & 19.19$^{+0.33}_{-0.25}$ & 0.13$^{+0.56}_{-0.11}$ &$\cdots$&$\cdots$& \\ \hline
\hline 
\end{tabular}
\end{center}
\end{table*}

\section{Conclusion}
We have analyzed high resolution X-ray absorption/emission data observed by
Chandra  and  Suzaku  to determine the physical properties of
the hot ISM
toward  the galactic bulge (4U 1820$-$303) direction with an estimate for the 
contribution from normal stars in the soft X-ray band. 

A two component plasma model can reproduce the absorption and emission spectra. 
One model assumes that only one component contributes to the absorption  in  front of 4U 1820$-$303.
  The temperature,  column  density and length of the front plasma are determined as 1.7$^{+0.2}_{-0.2}$
  $\times10^6$ K, 3.4$^{+2.1}_{-1.2} \times10^{19}$ cm$^{-2}$ and 3.1$^{+4.5}_{-1.8}$  kpc. 
  The temperature and emission measure of the back-side plasma are determined as 3.9$^{+0.4}_{-0.3}$ 
  $\times10^6$ K and 18.9$^{+5.1}_{-4.0}$ cm$^{-6}$ pc.
  This model is consistent with a scheme with  a hot X-ray bulge and an  exponential disk model 
  obtained from  extragalactic source observations. If there are two plasma components contributing to the absorption, 
   a thin warm plasma disk with temperature of $1.0^{+0.2}_{-0.2} \times10^{6} $ K and a length 
 of $0.13^{+0.56}_{-0.11} $ kpc is confined to the Galactic disk, and it is not pressure-balanced 
 with a hot thick disk with temperature of $3.2^{+0.2}_{-0.3} \times 10^{6}$ K.
 
\bigskip
Part of this work was financially supported by Grant-in-Aid for Scientific Research (Kakenhi) 
by MEXT, No. 20340041, 20340068, and 20840051. TH appreciates support from the JSPS 
research fellowship and the Global COE Program ``the Physical Sciences Frontier'', MEXT, Japan.


\end{document}